\documentclass[aps,epsfig,prb,twocolumn,longbibliography]{revtex4-2}

\usepackage{amsmath,amssymb,amsfonts,latexsym}
\usepackage{graphicx,xcolor}

\def\nn{\nonumber}

\newcommand{\sgn}{\mathop{\mathrm{sgn}}\nolimits}

\begin{document}

\title{Residual gauge theory for quanta of surface plasmons}

\author{Ken-ichi Sasaki}
\email{ke.sasaki@ntt.com}
\affiliation{NTT Basic Research Laboratories and NTT Research Center for Theoretical Quantum Information, NTT Corporation,
3-1 Morinosato Wakamiya, Atsugi, Kanagawa 243-0198, Japan}

\date{\today}

\begin{abstract}
 We develop a gauge-theoretical framework to investigate the quanta of surface plasmons.
 Our formulation, based on quantum electrodynamics,
 highlights the importance of residual gauge symmetry.
 We emphasize that residual gauge symmetry, which imposes constraint equations on physical states, 
 is fundamentally linked to Joule heating.
 This framework is applied to metals, semiconductors, and quantum Hall states, suggesting 
 the presence of a latent transverse electric mode and that 
 the quanta have the ability to maintain light-matter entanglement.
\end{abstract}

%\pacs{73.20.-r, 73.40.-c, 72.80.Vp}
\maketitle

The photon is often represented by a plane wave,~\cite{Weinberg1995,Peskin1995,Heitler1984}
but this is only one idealized solution among many to the wave equation
derived from Maxwell’s equations.
Even in free space, solutions with complex shapes exist.~\cite{Ranada1989}
For example, stable topological solutions to classical Maxwell’s equations,
whose electric and magnetic field lines
encode all torus knots and links, have been discovered in recent years.~\cite{Kedia2013}
Simpler than these,
the wave equation also allows for waves that are exponentially 
localized in the direction transverse to their propagation.
These localized waves, referred to as surface and evanescent waves,
have imaginary wavevectors and exhibit unusual frequencies
that are significantly lower than those of simple plane waves.~\cite{Polder1971}
This reduced frequency is intriguing because 
quantum coherence and discreteness often become more evident 
when the energy scale decreases relative to thermal fluctuations.
%This reduced frequency is intriguing because 
%quantum effects become more pronounced at lower energy.
However, a single exponentially localized function 
diverges in one spatial direction and is not normalizable.
To obtain a normalizable and stable photon state,
two localized functions must be appropriately bound together.

In this paper, we explicitly construct such a localized single-photon state 
by combining two localized waves via
a two-dimensional electron gas (2DEG), based on gauge symmetry in quantum electrodynamics (QED).
The classical counterpart of this quantum state manifests as a collective mode 
known as surface plasmons in classical electrodynamics.~\cite{Raether1988a,nakayama74}
Our focus is on their quantum aspects---the quanta of surface plasmons.
These quanta are not only of interest in basic research but may also have potential technological applications.
To transmit, receive, and store information carried by photons in free space, 
it is ultimately necessary to integrate a single photon into an electrical network.
A fundamental challenge in this endeavor is controlling a long-lived entangled state of the form
$|{\rm photon} \rangle + |{\rm matter}\rangle$ through the interaction 
between a single photon and matter.~\cite{Inoue2006,Rakonjac2021}

The collective motion of electrons in solids is intimately connected to their thermal and quantum behavior.
Bohm and Pines formulated a collective description of electrons interacting through Coulomb forces 
by decomposing the density fluctuations into collective and individual parts.~\cite{Bohm1951,Pines1952,Bohm1953}
The collective part behaves almost freely, while the individual part represents random thermal motion of electrons.
A similar conceptual framework can also be found in the theory developed by Rytov, Polder and Van Hove.
In their theory of radiative heat transfer between closely spaced bodies,~\cite{Polder1971} 
the current density is separated into an induced (collective) component and a random thermal source
with zero mean and finite correlation determined by the fluctuation-dissipation theorem.
Importantly, they accounted for not only propagating (far-field) 
but also (near-field) evanescent modes in thermal radiation
(for recent reviews, see~\cite{Joulain2005,Volokitin2007,Biehs2021a}). 
Although Polder and Van Hove did not explicitly refer to the work of Bohm and Pines, 
their formulation shares a similar structure. 
Both theories decompose fluctuations into collective (induced) and individual (random) components, 
although the statistical averaging is introduced at different stages of the formulation.
This is consistent with the continuity relation linking the charge density and current,
which naturally connects the density-based and current-based formulations of Bohm-Pines and Polder-Van Hove.
This structural analogy, despite their independent origins, 
suggests a deeper correspondence between collective electron dynamics and thermal electromagnetic fluctuations.
In the present work, 
we focus on the current density associated with collective modes that possess a finite lifetime,
which are described by the fluctuation-dissipation theorem as 
ultimate manifestations of the underlying quantum fluctuations in plasmonic excitations.

QED serves as the prototypical model of gauge theories, 
where gauge symmetry and its spontaneous breaking often dictate physical phenomena.~\cite{Weinberg1995,Peskin1995}
While QED has been employed in previous studies to investigate surface plasmons,~\cite{Archambault2010,Todorov2012,Hanson2015,Miwa2021,Rokaj2022}
the present formulation provides a framework in which 
residual gauge symmetry and boundary counterterms are treated explicitly.
Residual gauge transformation corresponds to a constant shift of the gauge field, 
and the requirement of invariance under this transformation 
imposes a constraint on physical states, which can be readily solved.
Interestingly, the residual gauge symmetry not only imposes constraints on physical states 
but also reveals a deep connection to thermodynamic concepts. 
For example, we demonstrate that 
this constraint equation emerges from a unitary transformation
of the gauge field, generated by an operator associated with Joule heating.
The Joule-heating operator represents the fundamental channel through which 
electromagnetic energy is transferred into the electronic degrees of freedom.
This irreversible energy exchange, even if weak, is intrinsic to the existence of plasmon quanta.
A perfectly lossless plasmon with an infinite lifetime can be defined theoretically,
but such a mode would be completely isolated, contradicting the very coupling that gives rise to it.
This suggests a fundamental connection between residual gauge symmetry
and Joule heating (or thermodynamics).
In a periodic space,
a residual gauge transformation corresponds to the Aharonov-Bohm effect, 
highlighting the physical reality of the gauge field.~\cite{Feynman1989,Olariu1985}
Therefore, our findings indicate an unexpected link
between spatial topology and the physics of surface plasmon quanta,
mediated by residual gauge symmetry.

Our formulation, based on a residual gauge transformation, 
bears a conceptual similarity to the Bohm-Pines formulation, in which 
a canonical transformation is 
introduced to extract the plasmon field.~\cite{Bohm1953}
In their formulation, employing the temporal gauge (with the scalar potential set to zero), 
the vector potential is decomposed into longitudinal and transverse components,
${\bf A} = {\bf A}_{\parallel} + {\bf A}_{\perp}$, where 
the longitudinal part is curl-free $\nabla \times {\bf A}_{\parallel}=0$ 
($\nabla \cdot {\bf A}_{\parallel}\ne 0$), 
and the transverse part is divergence-free 
$\nabla \cdot {\bf A}_{\perp}=0$ ($\nabla \times {\bf A}_{\perp} \ne 0$).
The plasmon field is constructed from the long-wavelength components of 
${\bf A}_{\parallel}$ and its canonical conjugate ${\bf E}_{\parallel}$,
with Gauss's law serving as a subsidiary condition that 
fixes the relationship between the electron density and ${\bf E}_{\parallel}$.
Through the canonical transformation, 
this subsidiary condition is converted into 
a new one ${\bf E}_{\parallel}|\Psi \rangle=0$, which imposes that
any longitudinal component not associated with the charge density must eliminate
the plasmon state.
%The Bohm-Pines formulation provides an example in which 
%the collective excitation modes can be viewed as an active, rather than passive, 
%gauge transformation---one that acquires physical significance and reality 
%through the emergence of collective behavior.
It should be noted, however, that there was active discussion concerning the subsidiary
condition: it might be too restrictive, 
leading to non-normalizable states, 
and should therefore be modified.~\cite{Adams1955,Kuper1956,Kanazawa1957,Bohm1957}

A surface plasmon quantum possesses two degrees of freedom, 
analogous to the polarization states of a single photon in free space.
These are referred to as transverse magnetic (TM) and transverse electric (TE) modes.~\cite{Raether1988a,Nakayama1985}
In contrast to the degenerate helicities of a free-space photon,
protected by spatial inversion symmetry,~\cite{Weinberg1995}
the TM and TE modes are non-degenerate due to the broken rotational symmetry about the propagation axis
introduced by the 2DEG matter.
Polarization states of photons serve as fundamental 
resources for encoding quantum information in optical telecommunications.
Similarly, the distinct TM and TE modes of surface plasmons are anticipated to play critical roles 
in quantum information processing within electrical networks. 
To investigate their behavior in different media, 
we explicitly calculate the mode structures under realistic conditions. 
Our results reveal that, in metals, the TE mode forms a zero-energy flat band, 
while in semiconductors with an intrinsic energy gap, the TE mode appears slightly below the gap energy. 
Furthermore, when an external magnetic field is applied, 
the Lorentz force induces coupling between the TM and TE modes, 
although this coupling remains negligible in the bulk. 
Notably, at the boundary of a two-dimensional electron gas (2DEG), 
the TM and TE modes hybridize and reconstruct into a chiral excitation known 
as the edge magnetoplasmon.~\cite{Mast1985,volkov88,Aleiner1994,Ashoori1992,Kumada2014a,Brasseur2017}

Surface plasmons are widely studied across physics, chemistry, 
and biology due to their fundamental link to visibility.~\cite{Raether1988a,nakayama74}
A familiar example is the air-noble metal interface, 
where the plasmon frequency is comparable to that of free-space photons. 
However, because of the high energy scale, this regime is not our focus. 
Instead, we study plasmons in two-dimensional electron gases (2DEGs), 
using layered materials such as graphene (metallic) and molybdenum disulfide (MoS$_2$, semiconducting) as examples. 
In these systems, the plasmon frequency is significantly reduced by confinement between localized photon modes, 
enabling access to various quantum phases realized at low temperatures, despite being invisible to the human eye.

This paper consists of three sections.
Section~\ref{sec:for} presents our gauge-theoretical formulation for constructing quanta of surface plasmons.
In Section~\ref{sec:app}, we apply this formulation to examine the properties of the quanta
in metals, semiconductors, and under an external magnetic field.
Section~\ref{sec:dis} provides the discussion.
Let us summarize the notations used in this paper.
The position vector is expressed as ${\bf r}=(x,y,z)$, and time is denoted by $t$.
Although the field is a function of space-time, we omit this dependence 
in most cases for notational simplicity.
The symbol $\nabla$ represents the nabla operator, 
and $\partial_t \equiv \partial/\partial t$ denotes the partial derivative with respect to time.
$c$ is the speed of light in vacuum, and $h$ ($\hbar$) is the (reduced) Planck constant.
The permittivity of free space is given by $\epsilon_0=1/\mu_0 c^2$,
which should be replaced by the permittivity of the relevant medium.

\section{Formulation}\label{sec:for}

Our treatment of surface plasmons is based on 
the canonical quantization of Maxwell's classical theory of electromagnetism.~\cite{Weinberg1995}
The fundamental knowledge required to understand Sections~\ref{sec:bc},~\ref{sec:quanta}, and~\ref{ssec:cs}
is provided in Section~\ref{ssec:bk}.

\subsection{Fundamental knowledge}\label{ssec:bk}

The fundamental variables in QED 
are the gauge fields consisting of $A_0$ and ${\bf A}$,
rather than the electric ${\bf E}$ and magnetic ${\bf B}$ fields in classical Maxwell's equations.
The observable ${\bf E}$ and ${\bf B}$ fields are expressed in terms of the gauge fields as 
${\bf E} = - \nabla A_0 - \partial_t {\bf A}$
and ${\bf B} = \nabla \times {\bf A}$.
Substituting this ${\bf E}$ into Gauss's law,
$\nabla \cdot {\bf E} = \rho/\epsilon_0$,
and imposing the Coulomb gauge condition, $\nabla \cdot {\bf A} = 0$,
we obtain $\nabla^2 A_0 = - \rho/\epsilon_0$.
Using the Green's function identity 
$\nabla^2 \left( 1/4\pi|{\bf r}-{\bf r'}| \right) = -\delta({\bf r}-{\bf r'})$,
we obtain the Coulomb potential:
$A_0({\bf r}) = \int_V d{\bf r'} (\rho({\bf r'})/4\pi \epsilon_0|{\bf r}-{\bf r'}|)$.
Thus, $A_0$ is not an independent field variable but a functional of matter fields.
The longitudinal component $A_0$ and the transverse components
${\bf A}$ are mathematically well separated by Coulomb gauge.
The transverse components of ${\bf E}$ act as the canonical conjugates of ${\bf A}$,
and we denote them as ${\boldsymbol \Pi}_\perp$ ($\equiv \partial_t {\bf A}$).
The equal-time commutator is constructed to satisfy the Coulomb gauge condition:
$[A_i({\bf r}),\Pi_{\perp j}({\bf r'})] = i(\hbar/\epsilon_0) \{ \delta_{ij} \delta({\bf r}-{\bf r'})
 + \frac{\partial}{\partial r_i} 
\frac{\partial}{\partial r_j} ( \frac{1}{4\pi|{\bf r}-{\bf r'}|}) \}$.

Because the energy density of electromagnetic fields is defined as 
$(\epsilon_0/2)({\bf E}^2 + c^2 {\bf B}^2) - {\bf j}\cdot {\bf A} + \frac{1}{2} \rho A_0$,
the Hamiltonian $H$ is given by $H_0 + V + V_{C}$,
where
\begin{align}
 & H_0 = \int_V \frac{\epsilon_0}{2}\left[ {\boldsymbol \Pi}_\perp^2 + c^2(\nabla \times {\bf A})^2 \right]d{\bf r}, \\
 & V = - \int_V {\bf j} \cdot {\bf A} d{\bf r}, \\
 & V_{C} = \frac{1}{2} \int_V \rho({\bf r}) A_0({\bf r})d{\bf r} = \frac{1}{2}
 \iint \frac{\rho({\bf r}) \rho({\bf r}')}{4\pi \epsilon_0|{\bf r}-{\bf r}'|} d{\bf r}d{\bf r}',
\end{align}
plus the Hamiltonian for matter fields $H_M$.~\cite{Weinberg1995}
Note that $A_0$ disappears from $H$, and we no longer need to consider it. 
Since the time evolution of the operators in the Heisenberg picture
is given by ${\bf A}({\bf r},t) = e^{iH t} {\bf A}({\bf r},0)e^{-iH t}$,
${\boldsymbol \Pi}_\perp({\bf r},t) = e^{iH t} {\boldsymbol \Pi}_\perp({\bf r},0)e^{-iH t}$,
and ${\bf j}({\bf r},t)= e^{iH t} {\bf j}({\bf r},0)e^{-iH t}$,
we obtain the wave equation:
\begin{align}
 (\partial_t^2 - c^2 \nabla^2 ) {\bf A}({\bf r},t) = {\bf j}({\bf r},t)/\epsilon_0.
 \label{eq:fe}
\end{align}
Thus, we can restrict our focus to the Hilbert space composed of solutions to the wave equation
in free space, where the source term ${\bf j}$ vanishes, 
and impose boundary conditions at ${\bf j}\ne0$ on the solutions.
This current ${\bf j}$ satisfies $\nabla \cdot {\bf j}=0$ to be consistent with the Coulomb gauge.
By decomposing ${\bf j}={\bf j}_\parallel + {\bf j}_\perp$, 
where $\nabla \cdot {\bf j}_\perp=0$ and $\nabla \times {\bf j}_\parallel = 0$, 
the symbol ${\bf j}({\bf r},t)$ in Eq.~(\ref{eq:fe}) refers to ${\bf j}_\perp({\bf r},t)$.
We assume that matter fields are always localized at a thin plane with a nearly perfect flat surface
so that ${\bf j}({\bf r},t)$ is nonzero only on that plane.

Our primary interest lies in the evanescent waves of ${\bf A}$
that are tightly bound to an atomically thin plane located at a constant $z$ (which we take as $z=0$).
%where charged carriers exist.
The solutions of Eq.~(\ref{eq:fe}) with ${\bf j}= 0$,
which propagate in the $x$-direction with wavevector $k_x$, 
while remaining localized in the $z$-direction with a localization length $\xi$, 
are given by
\begin{align}
 \left\{
 \partial_t^2 - c^2 (\partial_x^2 + \partial_y^2 + \partial_z^2) \right\}
 e^{- |z|/\xi} e^{i(k_x x-\omega t)} = 0.
 \label{eq:sol}
\end{align}
The frequency $\omega$ must satisfy the dispersion relation
\begin{align}
 \frac{\omega^2}{c^2} - k_x^2 + \frac{1}{\xi^2} = 0.
 \label{eq:dispersion}
\end{align}
The values of $\omega$ and $\xi$ can be determined as functions of $k_x$
by imposing boundary conditions at $z=0$, as discussed in the next subsection [Sec.~\ref{sec:bc}].

Because Eq.~(\ref{eq:dispersion}) itself provides important insights
into surface waves, some of them are considered here.
Firstly, when $k_x = 0$, localized waves with a real $\omega$ 
cannot exist.
This is because if such an oscillating solution existed, 
then $\xi$ would become a purely imaginary number
\begin{align}
 \frac{1}{\xi} = \pm i \frac{\omega}{c}.
\end{align}
This represents a photon plane wave with the wavevector $k_z$ ($\xi^{-1}=\pm ik_z$),
which is not localized.
A localized wave with a pure imaginary $\omega$ may exist as a relaxational solution.
Meanwhile, when $\xi=\infty$ in Eq.~(\ref{eq:dispersion}), 
the dispersion relation reduces to $\omega = ck_x$, 
which is the standard form of the light dispersion.~\footnote{
To simplify notation, we assume $k_x\ge 0$ in most cases
and use $k_x$ to represent its magnitude $|k_x|$.}
When $\omega$ exceeds the light dispersion ($\omega > ck_x$) for some (small) $k_x$,
a real solution for $\xi$ in Eq.~(\ref{eq:dispersion}) does not exist, 
implying that the waves cannot be localized.

Secondly, if low-energy modes satisfying $\omega \ll ck_x$ exist, 
the corresponding localization length 
$\xi$ must be close to the lateral wavelength $k_x^{-1}$, or more precisely, slightly longer than $k_x^{-1}$.
For example, assuming that the boundary condition allows for 
a slowly propagating surface wave with a linear dispersion $\omega=v k_x$ where $v\ll c$,
$\xi$ is determined as
$\frac{1}{\xi^2} = \left(1-\frac{v^2}{c^2}\right) k_x^2$.
Note also that this mode is stable against radiative decay 
because its dispersion lies below the light dispersion.
However, it can decay into the electronic excitations or phonons
unless $v$ falls below the Fermi velocity or sound velocity, respectively.~\cite{Sasaki2014}
Finally, an important case, which will later be shown to approximately hold in several systems,
is given by
\begin{align}
 \omega = \omega_0 + v k_x,
 \label{eq:dis_mp}
\end{align}
where $\omega_0$ represents the energy gap of the mode.
Since the wave is given by $e^{-|z|/\xi} e^{ik_x (x-vt)} e^{-i\omega_0 t}$,
a pulse composed of different $k_x$ modes 
becomes a function of $x-vt$, exhibiting free propagation.
Because $\omega_0$ lies above the light dispersion for small $k_x$,
the gauge field can radiate out of the 2DEG as light emission, 
causing the evanescent waves satisfying $ck_x < \omega_0$ to become unstable.

\begin{figure}[htbp]
 \begin{center}
  \includegraphics[scale=0.5]{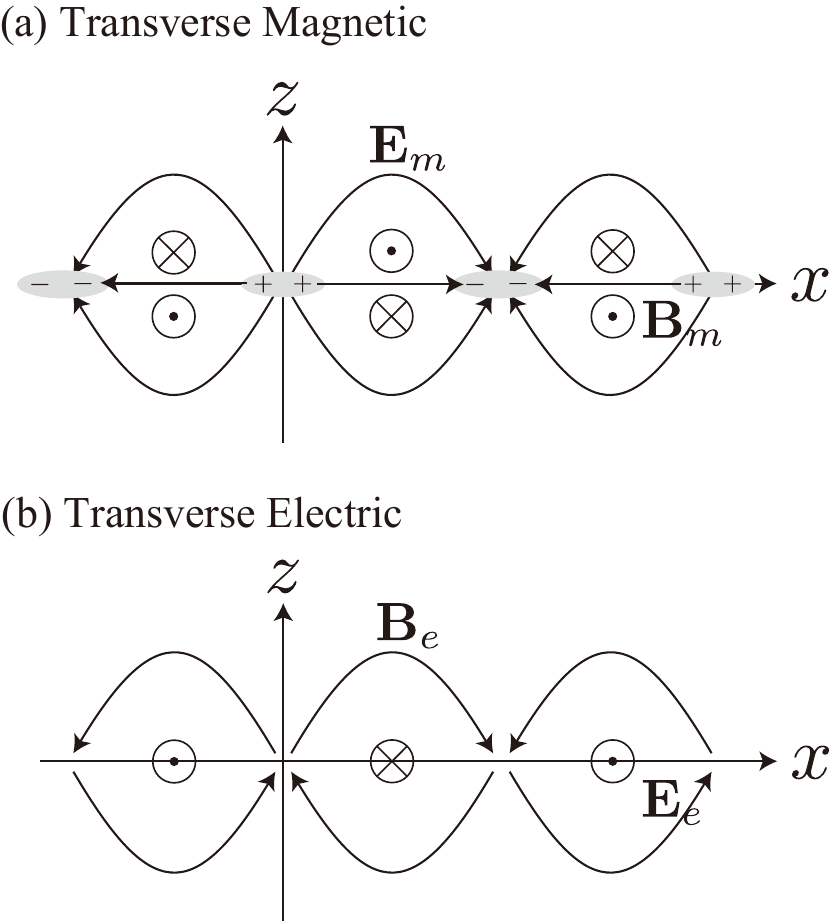}
 \end{center}
 \caption{
 (a) The transverse magnetic (TM) mode is defined as a surface wave with an electric field
 ${\bf E}_m=(E_x,0,E_z)$ and a magnetic field ${\bf B}_m=(0,B_y,0)$.
 The electric fields induce positive and negative charge densities, 
 which are associated with an electric current
 ${\bf j}=(j_x,0,0)$.
 According to Amp\`ere's circuital law,
 this current produces a discontinuity in $B_y$ at the plane $z=0$, 
 as indicated by the symbols $\otimes$ and $\odot$.
 (b) The transverse electric (TE) mode is defined as a surface wave with
 ${\bf E}_e=(0,E_y,0)$ and ${\bf B}_e=(B_x,0,B_z)$.
 According to Amp\`ere's circuital law,
 loops of the magnetic field induce electric current along the $y$-axis, ${\bf j}=(0,j_y,0)$,
 as indicated by the symbols $\otimes$ and $\odot$.
 We note that the electric (gauge) and magnetic fields are everywhere perpendicular both for the TM and TE modes, 
 ${\bf E}\cdot {\bf B}=0$ (${\bf A}\cdot {\bf B}=0$).~\cite{Ranada1989}
 }
 \label{fig:2}
\end{figure}

Using the surface waves given in Eq.~(\ref{eq:sol}),
we expand the gauge fields ${\bf A}$ 
with the creation and annihilation operators 
$a_\sigma^\dagger(k_x)$ and $a_\sigma(k_x)$ 
for a single photon with mode $\sigma$ ($\in\{m,e\}$) and wavevector $k_x$ as follows:
\begin{widetext}
\begin{align}
 & A_x(x,z,t) = \sum_{k_x} N_{m} e^{- \frac{|z|}{\xi}} \left(
 e^{i(k_x x-\omega t)} \epsilon_x(k_x) a_m(k_x) + e^{-i(k_x x-\omega t)} \epsilon_x^*(k_x) a^\dagger_m(k_x) \right) 
 = \sum_{k_x} A_x(x,z,t;k_x), \nn \\
 & A_y(x,z,t) = \sum_{k_x} N_{e} e^{- \frac{|z|}{\xi}} \left(
 e^{i(k_x x-\omega t)} \epsilon_y(k_x) a_e(k_x) + e^{-i(k_x x-\omega t)} \epsilon_y^*(k_x) a^\dagger_e(k_x) \right)
 = \sum_{k_x} A_y(x,z,t;k_x), \label{eq:Af}
 \\ 
 & A_z(x,z,t) = \sum_{k_x} N_{m} \sgn(z) e^{- \frac{|z|}{\xi}} \left(
 e^{i(k_x x-\omega t)} \epsilon_z(k_x) a_m(k_x) + e^{-i(k_x x-\omega t)} \epsilon_z^*(k_x) a^\dagger_m(k_x) \right)
 = \sum_{k_x} A_z(x,z,t;k_x), \nn 
\end{align}
\end{widetext}
where $N_{m}$, $N_e$ denote the normalization constants defined below,
$\epsilon_i$ ($i=\{x,y,z\}$) is the components of polarization vectors,
and ${\rm sgn}(z)$ is the sign function.
Since the wave modes propagate along the $x$-axis 
and are independent of the $y$-coordinate, 
the components $(A_x,0,A_z)$ and $(0,A_y,0)$ are decoupled in Coulomb gauge.
These two orthogonal modes are referred to as the transverse magnetic (TM) mode and 
the transverse electric (TE) mode, respectively.
As shown in Fig.~\ref{fig:2},
they originate from the 
two degrees of freedom associated with the orthogonal light polarizations $E_z$ and $E_y$
of the photon propagating in the $x$-direction.
We note that $A_x$ and $A_z$ are not independent 
due to the Coulomb gauge condition, 
$\partial_x A_x + \partial_z A_z = 0$,
which leads to
\begin{align}
 ik_x \epsilon_x(k_x) -\frac{1}{\xi} \epsilon_z(k_x) =0.
 \label{eq:CoulombG}
\end{align}
Thus, particularly for the TM mode, ``transverse'' does not refer to a transverse gauge field
but rather to the fact that only the ``magnetic'' field $B_y$
is transverse (or perpendicular) to the propagation direction.
Additionally, we note that $A_x$ and $A_y$ are continuous at $z=0$, 
whereas $A_z$ is discontinuous at this boundary.~\footnote{
Due to the discontinuity of $A_z$ at $z=0$,
the derivative $\partial_z A_z$ contains a delta function at $z=0$,
which means that the Coulomb gauge condition is not satisfied at $z=0$.
In reality, however, a 2DEG has a finite thickness ($2d$), and $A_z$ can be taken to be continuous.
For example, we may assume that $A_x = \cosh(z/\xi)e^{i(k_x x -\omega t)} \epsilon'_x$
and $A_z = -\sinh(z/\xi) e^{i(k_x x -\omega t)} \epsilon'_z$ for $z\in(-d,d)$.
By performing a gauge transformation with $A_{x,z} \to A_{x,z} + \partial_{x,z} \lambda(x,z)$ with
$\lambda(x,z)$ satisfying $(\partial_x^2 + \partial_z^2) \lambda = -g \cosh(z/\xi) e^{ik_x x}$
where $g=N_m \epsilon_z/(\xi \cosh(d/\xi) \sinh(d/\xi))$, 
the Coulomb gauge condition is satisfied everywhere.
}

%%%%%%%%%%%%%%%%%%%%%%%%%%%%
%%%%%%%%%%%%%%%%%%%%%%%%%%%%

The electromagnetic energy in space is given by the sum of 
the TM and TE modes as $H_{0} = H_m + H_e$,
where 
\begin{align}
 H_m
 &= \frac{\epsilon_0}{2} \int_V d{\bf r} (\Pi_{\perp,x}^2 + \Pi_{\perp,z}^2 + c^2B_y^2), \\
 H_e 
 &= \frac{\epsilon_0}{2} \int_V d{\bf r} (\Pi_{\perp,y}^2 + c^2 B_x^2 + c^2B_z^2).
\end{align}
Substituting Eq.~(\ref{eq:Af}) into the above equations 
and considering that
the creation and annihilation operators satisfy 
the standard commutation relations such as
$[a_\sigma(k_x) ,a^\dagger_{\sigma'}(k'_x)] = \delta_{\sigma\sigma'} \delta_{k_x,k'_x}$ 
and 
$[a_\sigma(k_x) ,a_{\sigma'}(k'_x)] = 0$ ($\sigma,\sigma'\in \{m,e\}$),
we obtain two independent harmonic oscillators
\begin{align}
 H_m &= \sum_{k_x} \hbar \omega \left( a_m^\dagger(k_x) a_m(k_x) + \frac{1}{2} \right), \\
 H_e &= \sum_{k_x} \hbar \omega \left( a_e^\dagger(k_x) a_e(k_x) + \frac{1}{2} \right).
\end{align}
Here we set $\epsilon_x(k_x) =1$ ($\epsilon_y(k_x) = 1$)
and choose the normalization constants as
\begin{align}
 N_{m}^2 = \frac{\hbar}{2\epsilon_0 S} \left( \frac{1}{\omega \xi (k_x\xi)^2} \right),  
 N_{e}^2 = \frac{\hbar}{2\epsilon_0 S} \left( \frac{\omega}{(ck_x)^2 \xi} \right),
 \label{eq:norm}
\end{align}
where $S$ denotes the area of the plane.
The ground state of the photon is expressed as $|0\rangle \equiv |0_m\rangle|0_e\rangle$,
where each ground state satisfies $a_m(k_x)|0_m\rangle =0$
and $a_e(k_x)|0_e\rangle =0$.
For later convenience, 
let us introduce the photon wave functions for each $k_x$ as
\begin{align}
 & f_{m}(x,z;k_x) \equiv N_{m}
 \epsilon_x
 \begin{pmatrix}
  1 \cr
  \sgn(z) ik_x \xi
 \end{pmatrix} 
 e^{-|z|/\xi}e^{i k_x x}, \\
 & f_{e}(x,z;k_x) \equiv
 N_{e} \epsilon_y 
 e^{-|z|/\xi}e^{ik_x x}.
\end{align}
The two components of $f_m$ represent the $x$- and $z$-components,
while $f_e$ consists of a single component representing the $y$-component.

\subsection{Boundary condition}\label{sec:bc}

The evanescent waves in the regions $z>0$ and $z<0$ are bound to a thin plane at $z=0$
by an electric current ${\bf j}(x)$.
Evanescent waves depend on the current, 
and conversely, the current also depends on the evanescent waves. 
This provides an example where the boundary conditions cannot be specified in a simple manner, 
unlike the case where the gauge field is taken to vanish at the surface of a metal.
To ensure consistency with the Hamiltonian,
the boundary condition at $z=0$ must be properly defined.
To achieve this, 
we extend the concept of the local Hamiltonian---originally introduced in Ref.~\onlinecite{Sasaki2024b},
for propagating waves with $k_x=0$---to evanescent waves with nonzero $k_x$.~\cite{Sasaki2024a}

The local Hamiltonian is defined by integrating $H_0$ over an infinitesimal region around the surface.
We do not include $V$ in the definition of the local Hamiltonian, 
because $V$, together with $H_M$, is incorporated into the covariant derivative to ensure gauge invariance.
From $\frac{\epsilon_0 c^2}{2}{\bf B}^2$ in $H_0$, we obtain, for example,
$\frac{\epsilon_0 c^2}{2}\int_{0_-}^{0_+} B_x^2 dz = \frac{\epsilon_0 c^2}{2}\int_{0_-}^{0_+} B_x(-\partial_z A_y) dz 
= \frac{\epsilon_0 c^2}{2}[-A_y B_x]_{0_-}^{0_+} + \frac{1}{2}\int_{0_-}^{0_+} A_y (j_y+\epsilon_0 \dot{E}_y) dz$,
where we have used integration by parts, in which
the surface term $[-A_y B_x]_{0_-}^{0_+}$ appears.
The surface term does not vanish because 
$B_x =-\partial_z A_y$ is discontinuous at $z=0$, thought $A_y$ is continuous there.
The last term is obtained because Amp\`ere's circuital law gives 
$\epsilon_0 c^2 \partial_z B_x=j_y + \epsilon_0 \dot{E}_y$.
The current term does not vanish at $z=0$, 
whereas the infinitesimal integral $\int_{0_-}^{0_+} \dot{E}_{y}A_y dz$ can be neglected,
since the integrand is continuous at $z=0$.
%~\footnote{Because $B_x^2 > 0$, the sign used to remove the absolute value from
%$|A_y|_{z=0} (\partial_z A_y|_{z=0_+}-\partial_z A_y|_{z=0_-}) |$ is not determined here.
%It can be determined based on consistency with 
%Amp\`ere's circuital law or by choosing the sign of the current operators.}
There is no contribution from $\frac{\epsilon_0}{2}{\boldsymbol \Pi}_\perp^2$ in $H_0$,
because $\Pi_{x,y}$ are continuous at $z=0$.
Although $\Pi_z$ is discontinuous there, its contribution can be eliminated by imposing the Coulomb gauge.
Assuming that the charge carriers move only on the plane,
${\bf j}=(J_x(x),J_y(x),0)\delta(z)$,
we obtain $\int_{0_-}^{0_+} (A_x j_x + A_y j_y)dz = A_x J_x + A_y J_y$.
By adding all contributions, 
we obtain the local Hamiltonian as an integral over surface as
\begin{align}
 H_{local} = \frac{\epsilon_0 c^2}{2} \int_S dxdy \left\{
 A_x(x) {\cal B}^c_x(x) + A_y(x) {\cal B}^c_y(x) \right\}.
 \label{eq:Hlocal}
\end{align}
Here, the boundary operators ${\cal B}^c_x(x)$ and ${\cal B}^c_y(x)$ are defined as 
\begin{align}
 {\cal B}^c_{x}(x) &\equiv B_y(x,z)|_{z=0_+} - B_y(x,z)|_{z=0_-}
 + \frac{J_x(x)}{\epsilon_0 c^2}, 
 \label{eq:Bx}
 \\
 {\cal B}^c_{y}(x) &\equiv B_x(x,z)|_{z=0_-} - B_x(x,z)|_{z=0_+}
 + \frac{J_y(x)}{\epsilon_0 c^2}. 
 \label{eq:By}
\end{align}

The energy must remain invariant under a gauge transformation,
given by $A_x \to A_x + \partial_x \lambda$
and $A_y \to A_y + \partial_y \lambda$.
Clearly, ${\cal B}^c_x(x)$ and ${\cal B}^c_y(x)$ are gauge-invariant for any $\lambda$ 
if $J_{x,y}(x)$ remains unchanged.
Since we use the Coulomb gauge condition, 
$\lambda$ must satisfy $(\partial^2_x+\partial_y^2) \lambda=0$,
which almost fixes the gauge. 
However, a residual gauge transformation remains, which is not entirely fixed by the Coulomb gauge,
given by $\lambda(x,y) = C_x x + C_y y$.
This results in a constant shift, or the zero mode of the gauge field, $A_i(x) \to A_i(x) + C_i$.
%~\footnote{
%Though $\nabla^2 A_0 = - \rho/\epsilon_0$ still holds, $A_0$ acquires a linear potential 
%by gauge transformation as $A_0 + \dot{C}_x x + \dot{C}_y y$.} 
%The generator of gauge transformation would be $\int_{\partial S} \dot{C}_i dr_i + \int_V d{\bf r} \rho({\bf r})/\epsilon_0$.
Please note that 
in $H_{local}$, the gauge field $A_i(x)$ appears directly
as a multiplier of ${\cal B}^c_i(x)$.
Under such a residual gauge transformation,
the local Hamiltonian transforms as 
$H_{local} \to H_{local} + \frac{\epsilon_0 c^2}{2} \int_S dxdy(C_x {\cal B}^c_x(x) + C_y {\cal B}^c_y(x))$.
To ensure that a residual gauge transformation does not alter the energy of physical states, 
any physical state $|\Phi_i \rangle$ must satisfy the condition 
\begin{align}
 \langle \Phi_j| {\cal B}^c_{x,y}(x)| \Phi_i\rangle=0. 
 \label{eq:const}
\end{align}
If this condition is not met, 
a residual gauge transformation could introduce an arbitrary amount of energy,
which is not observed in nature.

${\cal B}^c_{x,y}(x)$ bears a close resemblance to Amp\`ere's circuital law.
Amp\`ere's circuital law is expressed as 
$c^2 \nabla \times {\bf B} = \frac{\bf j}{\epsilon_0} + \dot{\bf E}$,
and integrating it over an infinitesimal region across the thin plane yields
$\int_{0_-}^{0_+} (\partial_z B_x - \frac{j_y}{\epsilon_0 c^2}) dz = 0$
and $\int_{0_-}^{0_+} ( \partial_z B_y + \frac{j_x}{\epsilon_0 c^2} ) dz = 0$.
The integral term $\int_{0_-}^{0_+} \dot{E}_{x,y}dz$ is neglected because $\dot{E}_{x,y}$
is continuous at $z=0$.
Thus, we reproduce Eq.~(\ref{eq:const}).
%In general, Gauss’s law is used to define physical states, 
%since it serves as the generator of gauge transformations. 
%The time derivative of the generator vanishes 
%when both Amp\`ere's law and the continuity equation are satisfied. 
%Therefore, our constraint based on Amp\`ere's law Eq.~(\ref{eq:const}) 
%ensures that physical states remain physical at later times. 

Our formulation can be naturally understood within
the action principle, starting from the standard QED Lagrangian,
${\cal L} = - \frac{1}{4} F_{\mu \nu} F^{\mu \nu} + A_\nu j^{\nu}$,
where $F_{\mu\nu}=\partial_\mu A_\nu- \partial_\nu A_\mu$~\cite{Weinberg1995}.
Under a general variation with respect to $A_\nu$, 
the change in ${\cal L}$ contains a total derivative (boundary) term as
$\delta {\cal L} = (\partial_\mu F^{\mu \nu} + j^\nu)\delta A_\nu - \partial_\mu (F^{\mu \nu} \delta A_{\nu})$.
We introduce a fictitious ``boundary'' that encloses the 2DEG located at $z=0$.
This boundary divides space into two regions: 
an interior region containing the 2DEG, and an exterior region
corresponding to free space without any current density.
Rather than attempting to solve the Maxwell equations in full detail throughout the whole space,
a more physically relevant question is what boundary conditions should be imposed at this boundary.
This viewpoint is analogous to electromagnetic problems in conductors, 
where Maxwell’s equations alone are insufficient unless supplemented by appropriate boundary conditions, 
such as the vanishing of the tangential electric field at a metallic surface.
Although the location of the boundary is in principle arbitrary, 
physical observables remain invariant, 
and the boundary serves as a device for encoding the effective response of the system.
We employ the identity
$- \frac{1}{4} F_{\mu \nu} F^{\mu \nu} = 
\frac{1}{2} A_\nu \partial_\mu F^{\mu \nu} - \frac{1}{2} \partial_\mu ( F^{\mu \nu}A_\nu )$.
The total derivative term cannot be discarded,
because it does not vanish ($F^{\mu \nu}\ne 0$ and $A_\nu \ne 0$) at the boundary and
gives a nontrivial contribution to the action.
To ensure that the Maxwell equations hold everywhere
and that the variational principle is well posed in the presence of a boundary,
one must add the explicit boundary counterterm
${\cal L}_{bc} = \partial_\mu ( F^{\mu \nu}A_\nu)$.
This boundary counterterm is not introduced merely to cancel the total derivative,
but to ensure the correct boundary conditions:~\cite{Gibbons1977}
in the interior region, arbitrary variations of $A_\nu$ are allowed,
whereas on the boundary, only gauge variations
($\delta F^{\mu \nu}=0$ but $\delta A_\nu \ne 0$) are physically admissible.
With this counterterm, the total Lagrangian becomes ${\cal L} + {\cal L}_{bc}=
\frac{1}{2} A_\nu \partial_\mu F^{\mu \nu} + \frac{1}{2} \partial_\mu ( F^{\mu \nu}A_\nu ) + A_\nu j^{\nu}$.
By combining half of the interaction term with the first term on the right-hand side, 
we define the bulk Lagrangian
$\frac{1}{2} A_\nu (\partial_\mu F^{\mu \nu} + j^\nu)$.
The corresponding action is an extremum under variations of $A_\nu$
that satisfy the Maxwell equations $\partial_\mu F^{\mu \nu} + j^\nu=0$.
The action remains stationary under variations of $A_\nu$ at the boundary,
i.e., residual gauge variations,
because they do not alter $\partial_\mu F^{\mu \nu} + j^\nu=0$.
The remaining half of the interaction term combines with the remaining half of the boundary
counterterm to form the ``boundary'' Lagrangian: $\frac{1}{2} {\cal L}_{bc} + \frac{1}{2}A_\nu j^{\nu}$.
When integrated across an infinitesimal region around $z=0$,
this becomes $\frac{1}{2} A_\nu ([F^{z\nu}]^{0_+}_{0_-} + J^\nu)$.
The surface integral of this expression leads to the local Hamiltonian given in Eq.~(\ref{eq:Hlocal}).
[This also provides another explanation for
why the derivative term $\dot{E}_i A_i$ at $z=0$ can be neglected.]
In this derivation, the role of the boundary counterterm is essential for maintaining consistency
of the variational procedure and for making the emergence of the constraint transparent,
although its interpretation as a Hamiltonian rather than an energy functional remains subtle.
Without the addition of ${\cal L}_{bc}$, 
the boundary contribution becomes
$\frac{1}{2} A_\nu (-[F^{z\nu}]^{0_+}_{0_-} + J^\nu)$, 
which cannot be made consistent with the Maxwell equations
$\partial_\mu F^{\mu \nu} + j^\nu=0$
obtained under ``arbitrary'' variations of $A_\nu$.
While the interior region is dominated by current-induced constraints, 
the exterior region satisfies the source-free Maxwell equations $\partial_\mu F^{\mu \nu} = 0$.
The nonvanishing boundary term in the exterior action shows that 
residual gauge transformations act nontrivially at the boundary and therefore represent physical degrees of freedom.
%by a general variation with respect to $A_\nu$, because $j^\nu=0$ there.]
%When the Heisenberg equation for ${\boldsymbol \Pi}_\perp$ is considered with resepct to $H_{local}$,
%Amp\`ere's law is reproduced (except for the factor 2)
%from the commutation relation between ${\boldsymbol \Pi}_\perp$ 
%and ${\bf A}$ as $-\dot{\Pi}_{\perp,i}(x,z) = c^2 {\cal B}^c_i(x) \delta(z)/2$. 
%However, only ${\cal B}^c_{x,y}(x)$ remains.
%The sign in front of $J_{x,y}(x)$ is $+$ here, while it is $-$ for ${\cal B}^c_{x,y}(x)$.
%If we assume that the gauge fields are not operators but c-number variables (according to Ehrenfest's theorem),
%then Eq.~(\ref{eq:Hlocal}), together with these conditions ${\cal A}^p_{x,y}(x)=0$, leads to
%$H_{local}=-\int_S dx dy A_i(x) J_i(x)$ which corresponds to $V$ in classical electrodynamics.
%This also implies that the constraint of Eq.~(\ref{eq:const}) no longer holds when $H_{local} \ne 0$.
%These results are consistent with Eqs.~(\ref{eq:Bx}), (\ref{eq:By}), and~(\ref{eq:const}).
%After eliminating the residual gauge degrees of freedom,

We can rewrite Eqs.~(\ref{eq:Bx}) and (\ref{eq:By}) 
in more compact forms.
Putting Eq.~(\ref{eq:Af}) into Eqs.~(\ref{eq:Bx}) and (\ref{eq:By}), and
using Eqs.~(\ref{eq:dispersion}), (\ref{eq:CoulombG}), and $\partial_y A_z(x,z)=0$,
we obtain simpler expressions of the boundary operators (for each $k_x$),
\begin{align}
 & {\cal B}^c_{x}(x;k_x) = \frac{2\omega^2 \xi}{c^2} A_x(x;k_x) + \frac{J_x(x;k_x)}{\epsilon_0 c^2}, 
 \label{eq:calBx} \\
 & {\cal B}^c_{y}(x;k_x) = -\frac{2}{\xi} A_y(x;k_x) + \frac{J_y(x;k_x)}{\epsilon_0 c^2}.
 \label{eq:calBy}
\end{align}

As a special case of a physical state,
we consider the ground state.
For simple systems such as metals and semiconductors, 
the ground state can be expressed
as a direct product of the photon vacuum and 
the matter vacuum state $|vac\rangle$.
In this case, we obtain the constraint equation:
\begin{align}
 \langle 0|\otimes \langle vac| {\cal B}^c_{x,y}(x)|\Phi_i \rangle=0.
 \label{eq:constvac}
\end{align}
As we will show later, this is useful for finding $|\Phi_i \rangle$.
In the following arguments, we mainly concern the simple ground state
given by the direct product of the photon vacuum and the matter vacuum, as in Eq.~(\ref{eq:constvac}).
However, such ground state is merely one possible ground state realized in nature.
For instance, 
the ground state of a superconductor cannot be simply described as a direct product of the Fock vacuum of photons
and the matter vacuum, as shown by the Meissner effects, 
in which the vacuum expectation value of the current operator itself behaves as the gauge field.
%Thus, a superconducting state would be described as a direct product of a coherent state of photons
%and the matter vacuum.
We examine more about quanta of surface plasmons in a superconductor elsewhere.
In Sec.~\ref{ssec:qhe} we discuss quantum Hall (QH) effect,
which is another example where the ground state has an energy gap
similar to a superconductor and possesses unique edge states.
The ground state is considered simply as $|0\rangle \otimes |vac\rangle$, while
the $|vac\rangle$ of QH states has very rich structure.

Equation~(\ref{eq:const}) can be used to show that 
the expectation value of $H_{local}$ with respect to any physical states
is zero, i.e. $\langle \Phi_j |H_{local}|\Phi_i \rangle=0$.
This conclusion is derived by inserting $1=\sum_i |\Phi_i \rangle \langle \Phi_i |$ 
consisting of all physical states
in the Hilbert space between $A_{x,y}(x)$ and ${\cal B}^c_{x,y}(x)$ and use Eq.~(\ref{eq:const}).
This zero-energy condition also validates the procedure in determining the normalization constants from $H_m$ and $H_e$
only.

However, there is an important exception of this zero-energy condition when the formulation 
is applied to a dynamical creation (emission) of a new photon
from a real excited state.
In this special case, only the terms $\frac{1}{2} \int_S dxdy(A_x J_x + A_y J_y)$
in $H_{local}$ contributes to
$\langle \Phi_j |H_{local}|\Phi_i \rangle$ and can make it to be nonzero.~\cite{Sasaki2024b}
An elaborate approach to handling this case is to make
a residual gauge transformation to acquire physical reality:
there is a special residual gauge transformation $C_j$
(in $A_j \to A_j + C_j$) that can be regarded as
an operator that changes $|vac\rangle$ into
a real excited state $|exc\rangle=\hat{C}_j|vac \rangle$.
%~\footnote{
%This treatment of the residual gauge is somewhat similar to the Anderson–Higgs mechanism.~\cite{Peskin1995} 
%However, the ground state is not a particular condensate, but rather a normal Fock vacuum.
%}
$C_j$ may depend on time $t$, as long as it does not depend on $x$ and $y$ 
(i.e., $\lambda(x,y,t)=C_x(t) x + C_y(t) y$).~\footnote{
In this case, $A_0\to A_0 - \dot{\bf C}\cdot {\bf r}$ in addition to $A_j \to A_j + C_j(t)$, 
and the electric field ${\bf E}$ (not ${\boldsymbol \Pi}_\perp$) caused by the residual gauge vanishes, of course.}
In this case, $C_j(t)$ becomes a dynamical variable, because it leads to a nonvanishing zero mode of 
$\Pi_{\perp,j}(t)$, which constitutes both $H_m$ and $H_e$.~\footnote{
This feature is reminiscent of QED in $1+1$ dimensions, namely, the Schwinger model.~\cite{Iso1990}}
In fact, according to Poynting's theorem, 
the continuity equation 
${\bf E}\cdot {\bf j} + \nabla \cdot {\bf S} + \dot{U}_{em} = 0$ holds,
where ${\bf S}$ represents the Poynting vector and 
$U_{em} = \frac{\epsilon_0}{2} ({\boldsymbol \Pi}_\perp^2 + c^2 {\bf B}^2)$
is the electromagnetic energy density.
Integrating this equation over an infinitesimal region around $z=0$ gives 
$-\epsilon_0 c^2 {\boldsymbol \Pi}_\perp(x)\cdot {\cal B}^c(x)+\int_{0_-}^{0_+} \dot{U}_{em} dz =0$.
Thus, when energy is indeed transferred from the electromagnetic field to the electronic system
(i.e., $\int_{0_-}^{0_+} \dot{U}_{em} dz\ne 0$),
we consider that $\langle 0 |\otimes \langle vac| \hat{C}_j {\cal B}^c_j(x) |\Phi_i \rangle$ is nonzero.
Without this thought, an excited state of matter, $|0\rangle \otimes |exc\rangle$, 
cannot itself be a physical state because Eq.~(\ref{eq:constvac}) gives 
$\langle vac| J_i(x) |exc \rangle =0$.
However, a physical excited state should have a nonzero $\langle vac| J_i(x) |exc \rangle$,
which is not in accordance to Eq.~(\ref{eq:constvac}) 
but compatible to $\langle 0 |\otimes \langle exc| {\cal B}^c_j(x) |\Phi_i \rangle \ne 0$
and $\langle 0 |\otimes \langle exc| V |\Phi_i \rangle \ne 0$.

In the present formulation,
the residual gauge mode is not a mere mathematical redundancy 
but represents a physical boundary degree of freedom 
that emerges when Maxwell’s equations are constrained by boundary conditions at the 2DEG interface. 
After Coulomb gauge fixing, the remaining transformation modifies the tangential electric field and thus couples to the surface current. 
Therefore, the residual gauge function $\lambda(x,y,t)$ 
acquires a dynamical meaning as a scalar field localized on the interface. 
As we will show in the next subsection, 
this field describes the current oscillation associated with the surface plasmon 
and provides the correct canonical variable that links the electromagnetic energy flow to Joule dissipation.

\subsection{Quanta of plasmon}\label{sec:quanta}

Finding physical states satisfying Eq.~(\ref{eq:constvac})
is straightforward.
We first define a single photon state 
with the wavevector $k_x$ as
$|m(k_x)\rangle =  a^\dagger_m(k_x)|0 \rangle$,
which satisfies 
\begin{align}
 \langle 0 | \Pi_{\perp,x}(x,z)| m(k_x)\rangle = (i\omega) f_{m,x}(x,z;k_x).
\end{align}
We then construct a quantum state of the TM mode as
\begin{align}
 & |{\rm TM}(k_x) \rangle 
 = |m(k_x)\rangle \otimes |vac\rangle + \nn \\
 &\int dx'
 \frac{\langle 0 |\Pi_{\perp,x}(x',0)|m(k_x)\rangle}{g}
 |0 \rangle \otimes J_x(x')|vac \rangle,
 \label{eq:TMW}
\end{align}
where $g$ is some constant (defined later in Eq.~(\ref{eq:sigmaxx})) introduced for dimensional consistency.
The first term on the right hand side represents the gauge field propagation alone 
$|m(k_x)\rangle$ without any excitation on the electronic system $|vac\rangle$, 
and the second term represents the excited electronic system $J_x(x')|vac \rangle$
without any excitation on the gauge field $|0 \rangle$.
This state expresses the superposition of a single photon 
and matter as a quantum entanglement.
Because directly comparing electronic states $J_x(x')|vac \rangle$ to measurements is challenging, 
it may be more effective to verify the quantum-mechanical discreteness of plasmons 
by using the quantum nature of evanescent photons $|m(k_x)\rangle$.

The physical state constraint Eq.~(\ref{eq:constvac}),
$\langle 0 | \otimes \langle vac | {\cal B}^c_{x}(x;k_x) |{\rm TM}(k_x) \rangle = 0$,
leads to the equation determining $\omega$: 
\begin{align}
 2i\omega \xi \epsilon_0 - \sigma_{xx}(\omega) = 0,
 \label{eq:TM}
\end{align}
where we have assumed that the current operators 
can be described within the local response regime at zero temperature:~\footnote{
To show $\langle {\rm TM}(k_x)|{\cal B}^c_x(x;k_x)|{\rm TM}(k_x)\rangle = 0$,
we need to assume $\langle vac | J_x(x)J_x(x')J_x(x'') | vac \rangle = 0$.}
\begin{align}
 & \langle vac | J_x(x) J_x(x') | vac \rangle = \sigma_{xx}(\omega) \delta(x-x') g, \label{eq:sigmaxx}
 \\ 
 & \langle vac | J_x(x) | vac \rangle = 0.
\end{align}
The dynamical conductivity $\sigma_{xx}(\omega)$ 
represents the contribution of the excitations with frequency $\omega$
that are excited by the application of $J_x(x')$ to $|vac \rangle$.~\footnote{In writing Eq.~(\ref{eq:sigmaxx}), we are assuming the operator product expansion: $J_x(x)J_x(x')\to F(x-x') J_x(x)$, where $F(x-x')$ is a singular c-number function.}
Due to entanglement, 
the normalization of the state deviates from unity, 
but it can be carried out by noting that
\begin{align}
 \langle {\rm TM}(k_x)|{\rm TM}(k_x)\rangle = 1 + \frac{\left( \int dx \right)}{g} 
 \omega^2 N_m^2 \sigma_{xx}(\omega).
\end{align}

A quantum state of the TE mode is obtained as
\begin{align}
 & |{\rm TE}(k_x) \rangle = |e(k_x)\rangle \otimes |vac\rangle + \nn \\
 &\int dx' \frac{\langle 0 |\Pi_{\perp,y}(x',0)|e(k_x)\rangle}{g} |0 \rangle \otimes J_y(x')|vac \rangle,
 \label{eq:TEW}
\end{align}
where
$|e(k_x)\rangle =  a^\dagger_e(k_x)|0 \rangle$
and $\langle 0 | \Pi_{\perp,y}(x,z) | e(k_x)\rangle = (i\omega) f_{e}(x,z;k_x)$.
Provided that $\langle vac | J_y(x) J_y(x') | vac \rangle = \sigma_{yy}(\omega) \delta(x-x') g$
and $ \langle vac | J_y(x) | vac \rangle = 0$,
the physical state constraint condition 
$\langle 0 | \otimes \langle vac | {\cal B}^c_{y}(x;k_x) |{\rm TE}(k_x) \rangle = 0$ leads to 
\begin{align}
 \frac{2}{i\omega \xi \mu_0} - \sigma_{yy}(\omega) = 0.
 \label{eq:TE}
\end{align}
We can know the dispersion relations of the TM and TE modes by
putting $\sigma_{xx}(\omega)$ and $\sigma_{yy}(\omega)$ into 
Eqs.~(\ref{eq:TM}) and (\ref{eq:TE}), respectively.
The dynamical conductivity can be derived using Kubo formula
from the Hamiltonian of matter and the calculated results depend on each system,
for example, they may depend on $k_x$ beyond the local response regime.
In Secs.~\ref{ssec:drude} and \ref{ssec:lorentz}, 
we consider simple models having the universal features 
applicable to metals and semiconductors, respectively.

The states Eqs.~(\ref{eq:TMW}) and (\ref{eq:TEW})
can be compactly rewritten using the operator
\begin{align}
 \hat{Q} \equiv - \int dx {\boldsymbol \Pi}_\perp(x) \cdot {\bf J}(x),
\end{align}
which yields
\begin{align}
 & |{\rm TM}(k_x) \rangle = \left( 1 - \frac{1}{g}|0\rangle \langle 0| \hat{Q} \right) |m(k_x) \rangle \otimes |vac \rangle, \\
 & |{\rm TE}(k_x) \rangle = \left( 1 - \frac{1}{g}|0\rangle \langle 0| \hat{Q} \right) |e(k_x) \rangle \otimes |vac \rangle.
\end{align}
These relations follow formally from the linear structure of Eqs.~(\ref{eq:TMW}) and (\ref{eq:TEW}).
The expectation values of $\hat{Q}$ in these physical states give
\begin{align}
 & \langle {\rm TM}(k_x) | \hat{Q} | {\rm TM}(k_x) \rangle 
 = \left( \int dx \right) \sigma_{xx}(\omega) \omega^2 N_{m}^2, \\
 & \langle {\rm TE}(k_x) | \hat{Q} | {\rm TE}(k_x) \rangle 
 = \left( \int dx \right) \sigma_{yy}(\omega) \omega^2 N_{e}^2.
\end{align}
They are direct consequences of the definitions and the physical state constraint, without additional assumptions.
The operator $\hat{Q}$ represents the quantum counterpart of Joule heating, 
since in classical electromagnetism the heat generation rate is given by $\int_V {\bf E}\cdot {\bf j}d{\bf r}$.~\footnote{Here, 
we assume that the surface terms $\int_{S} \nabla \cdot (A_0 {\bf j})$ 
and $\partial_t V_C $ both vanish so that there is no difference between $Q$ and $\hat{Q}$.}
Indeed, the real part of $\langle \hat{Q} \rangle$ given above
(${\rm Re}\langle \hat{Q} \rangle \propto {\rm Re}(\sigma_{ii})$)
corresponds to Joule heating, which causes a temperature rise in the 2DEG matter.
The temperature can become high due to strong localized modes with small $\xi$ values for the TM mode.
Since the normalization constants in Eq.~(\ref{eq:norm}) suggest then that $N_m^2 \gg N_e^2$,
Joule heating is primarily produced by the TM mode.
Note that the physical dimension of $g$ is the same as that of $\hat{Q}$.

This observation suggests that $\hat{Q}$ may generate photon-matter entanglement and
encode dissipation at the quantum level.
In particular,
\begin{align}
 e^{i\alpha \hat{Q}} A_j(x) e^{-i \alpha \hat{Q}} = A_j(x) - \frac{\hbar}{\epsilon_0} \alpha J_j(x),
 \label{eq:gaugeTr}
\end{align}
can be interpreted as indicating that the operator $\hat{Q}$ mixes the gauge field with the electronic current.
In fact, the boundary operators Eqs.~(\ref{eq:calBx}) and~(\ref{eq:calBy})
are obtained from $A_j(x)$ as 
\begin{align}
 {\cal B}^c_j(x;k_x) = 
 \frac{\mu_0 \epsilon_0}{\hbar \alpha_j}e^{i\alpha_j \hat{Q}} A_j(x;k_x) e^{-i\alpha_j \hat{Q}}, 
 \label{eq:gaugeTr2}
\end{align}
with an appropriate $\alpha_j$ 
($\alpha_x=-1/2\hbar \omega^2\xi$ and $\alpha_y=\xi /2\hbar c^2 $).
From this viewpoint, 
a plasmon quantum may be regarded not merely as a photon dressed with matter response, 
but rather as a new quantum object whose physical identity inherently involves both fields.

This interpretation resonates with Dirac's conjecture 
that the ``redundant'' gauge degrees of freedom might correspond to charged matter,
though his proposal was speculative.
%The $\hat{Q}$ operator, 
%in addition to generating the constraint equations, may also generate a special gauge transformation.
In gauge theories, 
the mathematics involve variables that encompass more degrees of freedom than are physically necessary,
much like a residual gauge transformation $A_j(x) \to A_j(x) + C_j$.
Dirac once hypothesized that 
electrons are exactly the surplus gauge degrees of freedom of light, 
and he attempted to derive laws that determine the electron's charge and mass.~\cite{PaulAdrienMauriceDirac1951,Schrodinger1952}
According to his formulation, the gauge field becomes proportional to the current.
His idea remains intriguing, 
and similar approaches are still 
under active investigation.~\cite{Hotta2001,He2014,Hawking2016,Hotta2016,casadio2024relaxation}
A plasmon quantum--an entangled state of an evanescent photon and a charged current--is neither purely light nor purely electron, 
and provides one example of the concept:
the right-hand side of Eq.~(\ref{eq:gaugeTr}) reflects 
Dirac's proposal that the mathematically redundant gauge degrees of freedom in QED ($\hat{C}_j$)
correspond exactly to electrons ($J_j$).
As we have mentioned, $\langle vac|J_j(x)|exc\rangle \ne 0$, where $|exc\rangle = \hat{C}_j|vac\rangle$.
If $\hat{C}_j$ is proportional to $J_j(x')$, then $\langle vac|J_j(x)|exc\rangle \ne 0$ implies that
$\langle vac|J_j(x)J_j(x')|vac\rangle \ne 0$. 
This condition is consistent with Eq.~(\ref{eq:sigmaxx}).
In the present context, the residual gauge mode appears not as a mathematical redundancy 
but as a physical surface degree of freedom localized at the 2DEG interface.
In this sense, residual gauge symmetry may acquire physical meaning through Joule dissipation and photon-matter hybridization.
Although this interpretation goes beyond what is strictly derived, it provides a conceptual bridge between surface plasmons, 
dissipation, and gauge symmetry, and may suggest a deeper correspondence between gauge structure and quantum matter.

As a simple application of Eq.~(\ref{eq:gaugeTr2}),
we can find a state that satisfies $\langle \Psi|A_j(x)|\Psi\rangle=0$
from any physical state $|\Phi\rangle$ just by multiplying the generator:
$|\Psi \rangle = e^{-i\alpha_j \hat{Q}} |\Phi \rangle$.
For example, since $|{\rm TM}(k_x)\rangle$ is an example of $|\Phi \rangle$,
the equality can be used to obtain a state with a vanishing expectation value of $A_j(x)$
as $|\Psi \rangle = e^{-i\alpha_j \hat{Q}} |{\rm TM}(k_x)\rangle$.
Similarly,
because $|0\rangle \otimes |exc\rangle$ 
is an instance of $|\Psi \rangle$,
the equality can be used to obtain a physical state of emission 
as $|\Phi \rangle = e^{i\alpha_j \hat{Q}} |0\rangle \otimes |exc\rangle$.
The state $|m(k_x)\rangle \otimes |vac\rangle$ is another example of $|\Psi \rangle$.

In general, 
Gauss’s law is used to define physical states,~\cite{Bohm1953}
since it serves as the generator of gauge transformations. 
The time derivative of the generator vanishes 
when both Amp\`ere's law and the continuity equation are satisfied. 
Therefore, our constraint based on Amp\`ere's law (Eq.~\ref{eq:const}) ensures 
that physical states remain physical at later times. 
Moreover, under the Coulomb gauge condition, 
the Joule heating operator $\hat{Q}$ commutes with Gauss’s law, 
ensuring that its action preserves gauge-invariant expectation values even for excited states. 
In this sense, the mode associated with $\hat{Q}$ 
corresponds to a physical dissipative excitation that remains consistent with 
Gauss’s law, which eliminates gauge-dependent longitudinal components 
carrying no independent dynamics.

\subsection{Coherent states}\label{ssec:cs}

Plasmons are found as classical solutions to Maxwell's equations.
While our primary interest lies in the quantum theory of plasmons,
it may be helpful to discuss the classical counterparts of these quantum states.
We define a coherent state using the normal ordering as:
\begin{align}
 |A \rangle = :e^{-i\int_V A({\bf r}) d{\bf r}}:|0\rangle, 
\end{align}
where $A({\bf r})$ represents either the TM or TE mode.
For the TM mode, it is written as
$A_m(x,z) = \sum_{k_x} f_m(x,z;k_x) a_m(k_x) + f^*_m(x,z;k_x) a_m^\dagger(k_x)$.
Normal ordering implies that all annihilation operators are placed on the right,
and all the creation operators are placed on the left.
Let us define $A \equiv \int_V A({\bf r}) d{\bf r}$. 
The field $A$ can be expressed as a sum of the 
creation $A^+$ and annihilation $A^-$ parts: $A=A^+ + A^-$.
Then, we have the expression $:e^{-iA}: = e^{-iA^+} e^{-i A^{-}} e^{\frac{[A^{+},A^{-}]}{2}}$.
As a result, 
$\langle 0 | A \rangle = e^{\frac{[A^{+},A^{-}]}{2}}$,
and $|A \rangle = e^{-iA^{+}} |0\rangle \langle 0 | A \rangle$.
The corresponding bra is defined as 
$\langle A | = \langle 0 | A \rangle \langle 0 | e^{+iA^{-}}$,
and the normalization is automatically established as
$\langle A |A \rangle = \langle 0 | e^{+iA^{-}}e^{-iA^{+}} |0\rangle \langle 0 | A \rangle^2 = \langle 0 | e^{[A^{-},A^{+}]} |0\rangle \langle 0 | A \rangle^2 = 1$.
With this notation, 
the coherent TM mode is defined as 
\begin{align}
 |\Phi_{A_m} \rangle &\equiv 
 \left( 1 - \frac{1}{2g}|A_m\rangle \langle A_m| \hat{Q} \right) |A_m \rangle \otimes |vac \rangle \nn \\
 &=|A_m \rangle \otimes |vac\rangle + \nn \\
 & \int dx' \frac{\langle A_m| \Pi_{\perp,x}(x',0) |A_m \rangle}{2g}  |A_m\rangle \otimes J_x(x') |vac \rangle.
\end{align}
From the constraint condition, we can obtain Eqs.~(\ref{eq:TM}) and (\ref{eq:TE})
for these coherent states, as well as those for the quanta.
In terms of the dispersion relation and related quantities such as propagation velocity, 
there appears to be no difference between the coherent state and the plasmon quantum.

The state $|\Phi_{A_m} \rangle$ is a simple direct product of 
a coherent state of light $|A_m\rangle$ and a matter state
$\left\{ 1+\frac{1}{2g} \int dx' \langle A_m| \Pi_{\perp,x}(x',0) |A_m \rangle J_x(x') \right\}|vac\rangle$.
This is in contrast to the entangled state of a plasmon quantum.
Moreover, coherent states can reproduce linear response relations through the expectation value:
\begin{align}
 \langle \Phi_{A_m}| J_x(x)|\Phi_{A_m} \rangle = \frac{\sigma_{xx}(\omega)}{2} \left(
 \langle A_m| \Pi_{\perp,x}(x,0) |A_m \rangle + h.c. \right). 
\end{align}
This result partly supports the existence of the dynamical conductivity described in Eq.~(\ref{eq:sigmaxx}).

\section{applications}\label{sec:app}

\subsection{Drude model}\label{ssec:drude}

\begin{figure}[htbp]
 \begin{center}
  \includegraphics[scale=0.5]{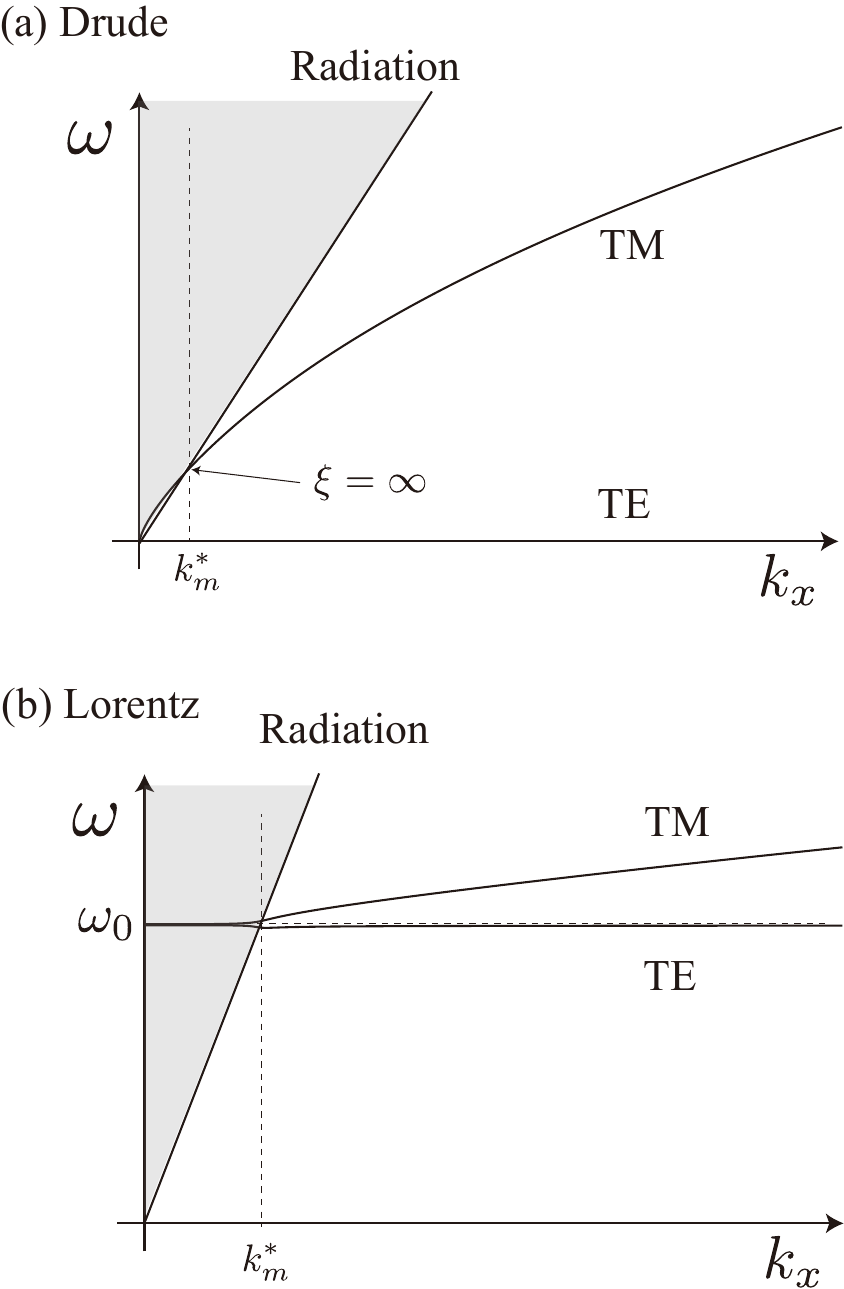}
 \end{center}
 \caption{
 (a) Drude model:
 The dispersion of the TM mode $\omega_p(k_x)$ is plotted as a function of $k_x$. 
 The intersection between the TM mode and the light dispersion $ck_x$
 occurs at $k_m^*$, where $\xi=\infty$. 
 Stable localized waves can exist for $k_x> k_m^*$. 
 The TE mode appears as a flat band along the $k_x$-axis.
 (b) Lorentz model:
 The dispersion of the TM mode appears as a nearly linear band with a constant group velocity.
 The TE mode exists below the gap and remains immobile.
 Although anticrossing is expected near the intersection of the TM (TE) and Radiation modes,
 we omit it here because our focus is on $k_x> k_m^*$.
 }
 \label{fig:1}
\end{figure}

The Drude model describes the motion of electrons in metals.
An electron is accelerated by electric fields, and 
its motion is governed by the classical equation of motion:
$m(d{\bf v}/dt+{\bf v}/\tau)=-e{\bf E}$, 
where $m$ is the effective mass of the electron, 
$\tau$ is the relaxation time, and ${\bf v}$ is the velocity.
Solving this equation and using the definition of the current density
${\bf j}\equiv -e n {\bf v}$ (where $n$ is the carrier density), 
we obtain the Drude model conductivities as:
\begin{align}
 \sigma_{xx}(\omega) = \sigma_{yy}(\omega) = \frac{\sigma_0}{1-i\omega \tau},
\end{align}
where $\sigma_0=ne^2\tau/m$ is the static conductivity.
In the collisionless case, 
where $\tau$ satisfies $\omega \tau \gg 1$,
we obtain from Eq.~(\ref{eq:TM}) the dispersion relation of the TM mode:~\cite{stern67,nakayama74}
\begin{align}
 \omega =\omega_p(k_x) - \frac{i}{2\tau}, \ {\rm where } \
 \omega_p(k_x) \equiv \sqrt{\frac{\sigma_0 k_x}{2\epsilon_0 \tau}}.
 \label{eq:omegap}
\end{align}
Note that in the denominator of $\omega_p(k_x)$, $\tau$ cancels out with $\tau$ in $\sigma_0$, 
making $\omega_p(k_x)$ independent of $\tau$ in the collisionless approximation.
Thus, $\tau$ is directly linked to the quantum fluctuations of the plasmonic excitation 
through the fluctuation–dissipation theorem.
Additionally, an exact calculation replaces $k_x$ with $\xi^{-1}$ in 
Eq.~(\ref{eq:omegap}).
Although we have intentionally replaced $\xi^{-1}$ with $k_x$,
this substitution requires careful consideration for small $k_x$ values [see Fig.~\ref{fig:1}(a)].
For example, 
the dispersion relation proportional to $\sqrt{k_x}$ may lead to a group velocity 
exceeding the speed of light for sufficiently small $k_x$.
This does not actually occur because of photons that are extended throughout the entire space. 
These photons exist in the shaded region above the dispersion labeled as ``Radiation'' in Fig.~\ref{fig:1}(a). 
In practice, such radiative states with small $k_x$ would smoothly transit from the TM mode.
On the other hand, for sufficiently large $k_x$,
the $\sqrt{k_x}$ dependence of the dispersion
%, which has been observed,~\cite{Yoshioka2024} 
is indeed reasonable, as the Poynting vector $S_x = \epsilon_0 c^2 (E_y B_z-E_z B_y)$, 
which represents the energy flow parallel to the 2DEG, becomes constant and independent of $k_x$: 
\begin{align}
 \epsilon_0 c^2\int_{-\infty}^{+\infty} dz
 \langle {\rm TM}(k_x)|\Pi_{\perp,z} B_y|{\rm TM}(k_x) \rangle
 = \frac{3\hbar}{2S} \frac{\omega_p(k_x)^2}{k_x}.
\end{align}
The factor of 3 originates from the fact that $S_x$ is proportional to $\{a_m(k_x),a_m^{\dagger}(k_x) \}$,
whose expectation value is 3 for the plasmon quantum.

It is also interesting to note that 
the Poynting vector, $S_z = \epsilon_0 c^2 (E_x B_y - E_y B_x)$, when spatially averaged for the TM mode, 
is given by
\begin{align}
 \bar{S}_z(z) &\equiv -\epsilon_0 c^2 \iint dxdy \Pi_{\perp,x} B_y \nn \\
 &= \sum_{k_x}\frac{\hbar}{2} \frac{i\omega^2}{(k_x \xi)^2}{\rm sgn}(z)e^{-\frac{2|z|}{\xi}}[a_m(k_x),a_m^{\dagger}(k_x)].
\end{align}
By assuming $k_x \xi \simeq 1$ and using Eq.~(\ref{eq:omegap}), we obtain 
\begin{align}
 & {\rm Re} [ \bar{S}_z(z) ] = \sum_{k_x}\frac{\hbar \omega_p(k_x)}{2\tau} {\rm sgn}(z)e^{-2k_x |z|},
\end{align}
which arises from the nonzero imaginary part of $\omega^2$.
The quantity ${\rm Re} [\bar{S}_z(z) ]$ vanishes when $\omega$ is purely real,
and this assumption is implicitly made when showing that 
the net thermal radiation of the evanescent modes into the vacuum vanishes.~\cite{Polder1971}
In classical electrodynamics, $\bar{S}_z(z)$ is proportional to the wavevector in the $z$ direction ($k_{z}$),
which is real (imaginary) for propagating (evanescent) waves.
The imaginary part of the Poynting vector represents an oscillation of electromagnetic energy rather than 
a net propagation, and it is non-vanishing for both TM and TE evanescent modes.
In the presence of a second body, the Poynting vector between two closely spaced bodies
is proportional to the wavevector inside the medium multiplied by its dielectric constant 
($\varepsilon^* k_{z}$ for TM and $k_z^*$ for TE).
The real part then becomes finite even for the evanescent modes.
This leads to the conclusion that energy transfer via evanescent waves requires the presence of a second body. 
In the case of quantum plasmons, the dissipation characterized by $\tau$ 
corresponds to the quantum fluctuation of the plasmons themselves.

The dispersion relation of the TE mode is obtained from Eq.~(\ref{eq:TE}) as
a purely relaxational mode:
\begin{align}
 \omega = - \frac{i}{\tau + \frac{\sigma_0 \mu_0}{2k_x}} 
 = - \frac{i}{\tau \left( 1 + \frac{n e^2}{2 \epsilon_0 m c^2 k_x} \right)}.
\end{align}
The correction to the lifetime by the term $\frac{\sigma_0 \mu_0}{2k_x}$
is negligible and practically unimportant.
However, the presence of $c$ suggests that the TE mode is closely related to light.
%As far as we know,the existence of this mode has not yet been explored experimentally.
The TE mode remains hidden in metals, but as we will show later, 
the TE mode itself begins to oscillate when electrons are bound to nucleons (Sec.~\ref{ssec:lorentz}). 
Moreover, in the presence of an external magnetic field,
the TE mode can hybridize to the TM mode at the boundary of 2DEG 
to form chiral edge magnetoplasmons (Sec.~\ref{ssec:emp}).

Polder and Van Hove showed that 
the heat transfer between closely spaced bodies is dominated by the TE mode
when the separation is on the order of a submicronmeter or less.~\cite{Polder1971}
More recently, Chapuis {\it et al.} have showed that, 
using both local and nonlocal models for a metal, 
this remains true even down to the nanometer scale.~\cite{Chapuis2008}
They argued that evanescent TE fields penetrate into the nearby body
and induce large eddy currents, which are dissipated by the Joule effect.
In other words, near-field radiative heat transfer can be viewed as nanoscale induction heating
at infrared frequencies.

In our treatment of plasmon quanta, 
the averaged Poynting vector for the TE mode is given by
\begin{align}
 \bar{S}_z(z) 
 = - \sum_{k_x}\frac{\hbar}{2} \frac{i\omega^2}{(k_x \xi)^2}{\rm sgn}(z)e^{-\frac{2|z|}{\xi}}[a_e(k_x),a_e^{\dagger}(k_x)].
\end{align}
The real part vanishes for a purely relaxational mode.
A nonvanishing heat flow occurs when the TE mode frequency acquires a finite real part,
as in the case of the TM mode.

%%%%%%%%%%%%%%%%%%%%%%%%%%%%
%%%%%%%%%%%%%%%%%%%%%%%%%%%%

\subsection{Lorentz model}\label{ssec:lorentz}

The Lorentz model describes an electron that is bounded to a nucleon, as in semiconductors.
The motion of the electron is governed by the classical equation of motion:
$m(d^2{\bf x}/dt^2+\omega_0^2 {\bf x} + \tau^{-1} d{\bf x}/dt)=-e{\bf E}$, 
where $\omega_0$ represents the energy band gap.
The dynamical conductivity is given by
\begin{align}
 \sigma_{xx}(\omega) = \sigma_{yy}(\omega)
 =\frac{\sigma_0}{1-i\left(\omega  - \frac{\omega_0^2}{\omega} \right) \tau}.
\end{align}
Near the resonance $\omega \sim \omega_0$, the electron can move within the crystal
and contribute to the current.
The Drude model can be obtained from the Lorentz model in the limit $\omega_0 \to 0$.

Substituting this $\sigma_{xx}(\omega)$ into Eq.~(\ref{eq:TM}),
we obtain the dispersion of the TM mode as
\begin{align}
 \omega = \sqrt{\omega_0^2 + \omega_p(\xi^{-1})^2} - \frac{i}{2\tau}.
\end{align} 
For small $k_x$, the dispersion is nearly flat at $\omega \sim \omega_0$.
For large $k_x$, the frequency can be approximated by the expansion
\begin{align}
 \omega \simeq \omega_0 \sqrt{1-\frac{(\sigma_0/\epsilon_0 c)}{4 \omega_0 \tau}}  + \frac{\omega_{p}(k_x)^2}{2\omega_0} + O(k_x^2),  
\end{align}
which takes the same form as the dispersion discussed in Eq.~(\ref{eq:dis_mp}).
The linear dispersion indicates that the TM mode propagates almost freely 
with a constant group velocity once it is excited [see Fig.~\ref{fig:1}(b)].~\cite{Nakayama1985}
The dispersion of the TE mode always appears below $\omega_0$ as
\begin{align}
 \omega = \frac{\omega_0}{\sqrt{1 + \frac{\sigma_0 \mu_0}{2\tau}\xi }} 
 - \frac{i}{2\tau \left( 1+  \frac{\sigma_0 \mu_0}{2\tau}\xi  \right)}.
\end{align}
The purely relaxational TE mode found in the Drude model is reproduced
by taking the limit $\omega_0 \to 0$ (apart from the factor 2 in the lifetime).
In order words,
the TE mode becomes oscillatory in semiconductors.
Unless $\frac{\sigma_0}{\epsilon_0 c} \frac{\xi}{c\tau}$ is unexpectedly large,
${\rm Re}[\omega]$ remains close to $\omega_0$,
and the TE mode has an almost negligible group velocity, making it effectively immobile.
Thus, it is reasonable to consider the TE mode as a collective mode of excitons (exciton-polaritons),
which are electron-hole bound states in a solid.~\cite{Nakayama1985}
The case where $ck_x \sim \omega_0$ or $\xi \sim \infty$
corresponds to a light emission from the annihilation (recombination) of an exciton.

When $\tau$ is sufficiently long,
the TE mode forms a nearly flat energy band with suppressed dissipation, 
appearing as a band aligned along the line of $\omega=\omega_0$
in Fig.~\ref{fig:1}(b). 
Owing to this flat energy band, 
the TE mode can undergo Bose-Einstein (polariton) condensation into a coherent state,
since a macroscopic numbers of modes accumulate near $\omega_0$.
We define the ground state $|S\rangle$ such that 
$\langle S|{\cal B}_y^c(x;k_x)|S \rangle=0$, which implies
$(2/\xi)  \langle A_y(x;k_x) \rangle_{S} = \mu_0 \langle J_y(x;k_x) \rangle_{S}$.
Substituting $\langle B_z(x) \rangle_{S} = \partial_x \langle A_y(x) \rangle_{S}$
into Amp\`ere's circuital law, $\partial_z B_x -\partial_x B_z = \mu_0 j_y $, we obtain
\begin{align}
 \frac{\omega^2}{c^2} \langle A_y(x) \rangle_{S} = \frac{2}{\xi d}\langle A_y(x) \rangle_S,
\end{align}
where $d$ is the effective thickness of the layer.
When the length $\sqrt{2\xi d}$ becomes comparable to $\lambda_0/\pi$ (where $\omega_0 = 2\pi c/\lambda_0$),
a nonzero solution $\langle A_y(x) \rangle_{S}$ emerges,
indicating a coherent, polariton-like condensate.

\subsection{Bulk Magnetoplasmon}

When a 2DEG is placed in an external static magnetic field,
the electrons experience the Lorentz force, causing
the current operators $J_x$ and $J_y$ to become mutually dependent.
As a result, the dynamical conductivity acquires a nonzero off-diagonal component,
given by
$\langle vac | J_x(x) J_y(x') | vac \rangle = \sigma_{xy}(\omega) \delta(x-x') g$.
A general physical state can be expressed as a superposition of the TM and TE modes:
\begin{align}
 |{\rm H} \rangle \equiv C_{m} |{\rm TM} \rangle + C_{e} |{\rm TE} \rangle.
\end{align}
The coefficients $C_m$ and $C_e$ are determined by the constraint conditions
$\langle 0 | \otimes \langle vac | {\cal B}^c_{x}(x) |{\rm H} \rangle = 0$
and $\langle 0 | \otimes \langle vac | {\cal B}^c_{y}(x) |{\rm H} \rangle = 0$,
which take the form of the following $2\times 2$ matrix equation:
\begin{align}
 \begin{pmatrix}
  2i \omega \xi \epsilon_0 - \sigma_{xx} & - \sigma_{xy} \cr
  -\sigma_{yx} & \frac{2}{i\omega \xi \mu_0} - \sigma_{yy}
 \end{pmatrix}
 \begin{pmatrix}
  C_{m} N_m \cr C_{e} N_e
 \end{pmatrix}
 = 0.
 \label{eq:22mat}
\end{align}

We assume that 
the motion of an electron in a static applied magnetic field, ${\bf B}_{a}=(0,0,B_{az})$,
is governed by the classical equation of motion:
$m(d{\bf v}/dt+{\bf v}/\tau)=-e({\bf E}+{\bf v}\times {\bf B}_a)$.
The corresponding conductivities are given by
\begin{align}
 \begin{split}
  & \sigma_{xx} = \sigma_{yy}= \frac{(1-i\omega \tau)\sigma_0}{(1-i\omega \tau)^2 +
  (\omega_c \tau)^2}, \\
  & \sigma_{xy} = -\sigma_{yx} = -\frac{(\omega_c \tau)\sigma_0}{(1-i\omega \tau)^2 +
  (\omega_c \tau)^2},
 \end{split}
\end{align}
where $\omega_c =eB_{az}/m$ is the cyclotron frequency.
The system supports three eigenmodes because, 
by setting the determinant of the matrix in Eq.~(\ref{eq:22mat}) to zero, 
we obtain a cubic equation for $\omega$:
\begin{align}
 & \frac{4\epsilon_0}{\mu_0} \left\{ (1-i\omega \tau)^2 + (\omega_c \tau)^2
 \right\} \nn \\
 & -(1-i\omega \tau) \sigma_0 \left( 2i\omega \xi \epsilon_0 +
 \frac{2}{i\omega \xi \mu_0} \right) + \sigma_0^2 = 0.
 \label{eq:deter}
\end{align}
When $\tau$ is sufficiently long,
the terms proportional to $\tau^2$ dominate.
By setting the coefficient of $\tau^2$ in Eq.~(\ref{eq:deter}) to zero,
we obtain the solutions:
\begin{align}
 \omega_{mp} = \pm \sqrt{\omega_c^2 + \omega_p(k_x)^2} - \frac{i}{\tau}.
 \label{eq:Dmp}
\end{align}
These are known as magnetoplasmons.~\cite{theis80,Ando1978,Pinczuk1993,Kushwaha2001}
Since $\omega_p(k_x)^2$ is linear in $k_x$,
the dispersion can be easily obtained from the expansion:
\begin{align}
 \omega_{mp} = \omega_c + \frac{\omega_p(k_x)^2}{2\omega_c} + O(k_x^2).
 \label{eq:Dmpsimp}
\end{align}
This corresponds to Eq.~(\ref{eq:dis_mp}), $\omega_{mp} = \omega_c + v k_x$, 
which is valid for small $k_x$.
The group velocity is given by
\begin{align}
 v = \frac{n}{eB_{az}} \frac{e^2}{4\epsilon_0}.
 \label{eq:gvelqhe}
\end{align}
Note that there is a critical wavevector below which 
the magnetoplasmon cannot exist, similar to the case of the Lorentz model.
For $k_x < k_x^*$, satisfying $\omega_c = c k^*_x$, 
the magnetoplasmon becomes unstable, leading to Landau emission with $\xi = \infty$.
For $k_x > k_x^*$, the magnetoplasmon is stable and 
primarily consists of the TM mode (i.e., the contribution of the TE mode is negligible $|C_m|\gg |C_e|$).
On the other hand, when $k_x < k_x^*$, the TE mode becomes dominant, corresponding to Landau emission.

The remaining solution of Eq.~(\ref{eq:deter})
is obtained under the assumption 
that the magnitude of $\omega$ is sufficiently small compared with $\omega_c$
and that $\tau$ is sufficiently long.
In this case, Eq.~(\ref{eq:deter}) can be approximated as
\begin{align}
 \frac{4\epsilon_0}{\mu_0} (\omega_c \tau)^2 
 -(1-i\omega \tau) \sigma_0 \left( \frac{2}{i\omega \xi \mu_0} \right) + \sigma_0^2 = 0,
 \label{eq:decayE}
\end{align}
which gives a purely relaxational state in the
presence of an external magnetic field,
\begin{align}
 \omega_{decay} = - \frac{i}{\tau + \frac{\omega_c^2}{\omega_p^2}\tau + \frac{\mu_0 \sigma_0}{2k_x}}.
 \label{eq:decaylifetime}
\end{align}
This mode consists of the TE mode:
$|C_e|\gg |C_m|$ holds. 
Meanwhile $|N_e|\ll |N_m|$ is satisfied, and $|C_m N_m| \gg |C_e N_e|$ still holds for the TE mode.
Although a purely relaxational state is not particularly prominent,
it plays an important role when considering edge magnetoplasmons near the boundary,
as shown in the next subsection.

\subsection{Edge Magnetoplasmon}\label{ssec:emp}

We have been assuming that the gauge fields are independent of the $y$ coordinate.
Here, we seek localized solutions that behave as $e^{-y/\xi}$, 
effectively modeling the edge states near the boundary $y=0$ of the 2DEG bulk ($y>0$).
The localization length in the $y$ direction does not necessarily 
have to match the localization length in the $z$ direction,~\cite{Sasaki2016}
but they do need to be scaled with each other.

Let us define new fields by multiplying 
$e^{-y/\xi}$ with each $k_x$ component of $A_{x,y,z}(x,z,t)$ in Eq.~(\ref{eq:Af}) and $(J_x,J_y)$
as
\begin{align}
 & \tilde{A}_i(x,y,z,t;k_x) = e^{-\frac{y}{\xi}} A_i(x,z,t;k_x), \\
 & \tilde{J}_i(x,y) = e^{-\frac{y}{\xi}} J_i(x).
 \label{eq:tildeJ}
\end{align}
Due to the Coulomb gauge condition for the deformed fields, $\nabla \cdot \tilde{\bf A}=0$, 
Eq.~(\ref{eq:CoulombG}) is modified to a new constraint because of
the non-vanishing $\partial_y \tilde{A}_y$, given by
\begin{align}
 \left\{
 ik_x \epsilon_x(k_x) - \frac{\epsilon_z(k_x)}{\xi} \right\} 
 N_m a_m(k_x) - \frac{\epsilon_y(k_x)}{\xi} N_e a_e(k_x)  = 0.
\end{align}
The gauge fixing imposes a constraint that enforces a coupling between the TM and TE modes,
resulting in physical states that become hybrids (entangled) of the TM and TE modes.
For example, a single-photon state is written as
$c_m|1_m\rangle |0_e\rangle + c_e |0_m\rangle|1_e \rangle$, where the coefficients $c_m$ and $c_e$ satisfy
Coulomb gauge $( ik_x \epsilon_x - \frac{\epsilon_z}{\xi} ) N_m c_m - \frac{\epsilon_y}{\xi} N_e c_e = 0$
and normalization $|c_m|^2+|c_e|^2=1$.
We can set $N_m=N_e(=N)$
by assuming that normalization is determined by $\epsilon_i$:
we can conclude that $\epsilon_z$ is modified
from $\epsilon_z=ik_x \xi$ [Eq.~(\ref{eq:CoulombG})] unless either $c_m$ or $c_e$ vanishes.
%In a more compact form, this can be expressed as $ik_x A^{-}_x - A_y^{-}/\xi - A_z^{-}/\xi = 0$, 
Moreover, the boundary operator ${\cal B}^c_y(x;k_x)$ in Eq.~(\ref{eq:calBy}) is modified because 
the magnetic field $B_x= \partial_y \tilde{A}_z - \partial_z \tilde{A}_y$ 
changes $B_x(x,z)|_{z=0_+}$ in Eq.~(\ref{eq:By}) 
%from $A_y^{-}/\xi$ to $e^{-y/\xi} (2A_y^{-}/\xi -ik_x A_x^{-})$
due to $\partial_y \tilde{A}_z \ne 0$ and the Coulomb gauge condition.
%where $A^-$ is the annihilation part of the field $A$.
Meanwhile, $B_y= \partial_z \tilde{A}_x - \partial_x \tilde{A}_z$ in Eq.~(\ref{eq:Bx}) remains unchanged
except for the scaling factor $e^{-y/\xi}$.
Thus, the constraint condition 
$e^{-y/\xi} \langle 0 | \otimes \langle vac | {\cal B}^c_{x}(x) |{\rm H} \rangle = 0$
is essentially unmodified.
We obtain a generalized version of Eq.~(\ref{eq:22mat}) 
with a modified 2$\times$2 matrix of the form: 
%\begin{align}
% \begin{pmatrix}
%  2 i\omega \epsilon_0 \xi - \sigma_{xx} - \frac{2}{i\omega \mu_0 \xi} 
%  & -\frac{2k_x}{\omega \mu_0} - \sigma_{xy} \cr
%  -\frac{2k_x}{\omega \mu_0} + \sigma_{xy}
%  & \frac{4}{i\omega \xi \mu_0} - \sigma_{xx}
% \end{pmatrix}
% \begin{pmatrix}
%  C_m \cr C_e
% \end{pmatrix}=0.
% \label{eq:mas_wa}
%\end{align}
%Here we consider the internal magnetic field perpendicular to the plane
%$B_z=\partial_x A_y - \partial_y A_x$.
%We assume that an internal magnetic field $B_z$ is suppressed, namely, 
%\begin{align}
% \frac{\xi^{-1} C_m +ik_x C_e}{i\omega}=0,
% \label{eq:constraint}
%\end{align}
%By using Eq.~(\ref{eq:constraint}), 
%we have the equations for the amplitudes as follows: 
\begin{align}
 \begin{pmatrix}
  2 i\omega \xi \epsilon_0 - \sigma_{xx}
  & - \sigma_{xy} \cr
  -\frac{2k_x}{\omega \mu_0} - \sigma_{yx}
  & \frac{4}{i\omega \xi \mu_0} - \sigma_{xx}
 \end{pmatrix}
 \begin{pmatrix}
  C_m N_m \cr C_e N_e
 \end{pmatrix}=0.
 \label{eq:mat4}
\end{align}

The vanishing determinant of the $2\times2$
matrix in Eq.~(\ref{eq:mat4}) leads to a cubic equation for $\omega$.
Since the corrections to Eq.~(\ref{eq:22mat}) are proportional to $\omega^{-1}$,
the dispersion of the magnetoplasmon remains unchanged.
Meanwhile, the purely relaxational mode acquires a nonzero real part in its frequency.
When $\omega \ll \omega_c$, assuming that $\tau$ is sufficiently long, 
the secular equation can be approximated as
\begin{align}
 & \frac{8\epsilon_0}{\mu_0} (\omega_c \tau)^2 
 -(1-i\omega \tau) \sigma_0 \left( \frac{4}{i\omega \xi \mu_0} \right) + \sigma_0^2 \nn \\
 &+ (\omega_c \tau) \sigma_0 \frac{2k_x}{\omega \mu_0} = 0.
\end{align}
Note that the first three terms have the same structure as those in Eq.~(\ref{eq:decayE}).
Under a strong magnetic field, 
the real part of $\omega$ is mainly determined by the first and fourth terms.
We obtain a solution with the linear dispersion:
\begin{align}
 \omega_{emp} = - \frac{\omega_p^2}{2\omega_c}
 -\frac{i}{\tau + \frac{\omega_c^2}{\omega_p^2} \tau + \frac{\mu_0\sigma_0}{4k_x}}.
 \label{eq:dischiral}
\end{align}
The dispersion acquires a non-zero real part, which is linear in $k_x$.
The fact that ${\rm Re}(\omega_{emp}) = -vk_x$ with $v \propto 1/B_{az}$ indicates that this
mode is chiral, meaning that the propagation direction
along the edge depends on the sign of $B_{az}$.
Additionally, the group velocity is suppressed by increasing $B_{az}$.
The negative sign in front of $\omega_p^2/2\omega_c$ corresponds to the negative exponent of $e^{-y/\xi}$.
When we start from $e^{+y/\xi}$, a positive sign appears in front of $\omega_p^2/2\omega_c$. 
All these properties of the solution are consistent with those
of edge magnetoplasmons.~\cite{volkov88,Wen1990a}
Thus, we identify this mode as the edge magnetoplasmon.
The null state of the modified matrix is found to satisfy $|C_mN_m|\sim |C_eN_e|$,
which proves that with $N_m=N_e$, the edge magnetoplasmons are hybrids (entangled) of the TM and TE modes.

It is worth noting that the intrinsic decay time of an edge magnetoplasmon
can be derived from the imaginary part of Eq.~(\ref{eq:dischiral}) as
\begin{align}
 \tau_{emp} \simeq \frac{\omega_c^2+\omega_p^2}{\omega_p^2}\tau,
 \label{eq:tauemp}
\end{align}
where $\mu_0\sigma_0/4k_x$ is omitted because it is small, on the order of 
$(\pi r_e^* n/k_x)\tau$, with $r_e^*$ being the solid-state analog of the classical electron radius, 
where the electron mass is replaced with the effective electron mass.
Because $\tau$ corresponds to the lifetime of magnetoplasmons (Eq.~(\ref{eq:Dmp})),
Eq.~(\ref{eq:tauemp}) shows that 
the ratio $\tau_{emp}/\tau_{mp}$ increases as $|B_{az}|$ increases.~\cite{Sasaki2016}
This result is consistent with the experimental findings reported in Ref.~\onlinecite{Yan2012}.

Here, we draw parallels and comment on the connection to the main discovery of edge magnetoplasmons. 
Volkov and Mikhailov first derived the dispersion relation of edge magnetoplasmons 
by solving an integral equation for the scalar potential using the Wiener-Hopf method.~\cite{volkov88} 
They assumed a steep confining potential at the sample edge and demonstrated that 
the resulting mode propagates in a direction determined by the external magnetic field and 
travels almost freely due to its nearly linear dispersion. 
Aleiner and Glazman extended the analysis to more general edge-potential profiles and 
showed that multiple plasmon modes with distinct group velocities can exist.~\cite{Aleiner1994}
They employed a Fredholm-type integral equation (derived from a linearized Euler equation) to obtain these modes, 
thereby elucidating the overall structure of edge magnetoplasmons, 
including those originally found by Volkov and Mikhailov.

In previous analyses, the internal magnetic field was usually neglected.
This simplification, however, makes it impossible to determine the lifetime of edge magnetoplasmons. 
Reference~\cite{Sasaki2016} showed that the intrinsic lifetime of the edge magnetoplasmon
is longer than that of bulk magnetoplasmon. 
That analysis was performed within the framework of classical electrodynamics, 
and in the present paper we obtain the same conclusion for plasmon quanta within the framework of QED (Eq.~\ref{eq:tauemp}). 
Our analysis rests on two key observations. 
First, a purely relaxational mode with a long lifetime exists inside a 2DEG (Eq.~\ref{eq:decaylifetime}). 
Second, this mode acquires a finite real part of its frequency through localization and begins to propagate (Eq.~\ref{eq:dischiral}). 
By showing that the properties of this localized mode are consistent with those of edge magnetoplasmons, 
one can identify the purely relaxational mode as the bulk counterpart of the edge magnetoplasmon 
and thereby determine its lifetime. 
The results further indicate that the internal magnetic field normal to the layer is strongly suppressed 
in the bulk of the 2DEG, which partly justifies the assumptions used in earlier theories and 
may lead to a more complete description of edge magnetoplasmons. 
Moreover, our QED-based analysis in the Coulomb gauge reveals a new feature: 
the quantum state of the edge magnetoplasmon is an entangled state consisting of the TM and TE modes.

\subsection{Quantum Hall Effect}\label{ssec:qhe}

The quantum Hall effect (QHE) is a phenomenon 
observed in a 2DEG at low temperatures and 
under the influence of a strong magnetic field.~\cite{Prange1990,Yoshioka2002,Ezawa2013,VonKlitzing2020}
The longitudinal resistance approaches zero, and the Hall conductance becomes quantized.
Unlike the ground states of ordinary metals and semiconductors, denoted by $|vac\rangle$,
the ground state in the QHE is unique because it exhibits different quantum phases 
depending on the filling factor $\nu = nh/eB$,
which is the ratio of the electron density $n$ to the magnetic flux quantum density $B/(h/e)$.
Thus, the ground state is expressed as a function of $\nu$ as $|{\rm QH}(\nu)\rangle$.
The following simple expression for the quanta of surface plasmons 
will be useful in emphasizing the dependence of $\nu$ on the entanglement:
\begin{align}
 a^\dagger|0\rangle \otimes |{\rm QH}(\nu) \rangle + |0\rangle \otimes J|{\rm QH}(\nu) \rangle.
\end{align}
%The correlation functions, such as $\langle {\rm QH}(\nu)|JJ| {\rm QH}(\nu) \rangle$, 
%differ from the dynamical conductivities discussed thus far.
Even for a fixed $\nu$,
there are multiple candidates for the ground state. 
Surface plasmons may provide a method 
to determine which candidate is valid through entanglement with light.~\cite{Du2019}
Moreover, the QHE is the first example
of a topological insulator, characterized by a bulk with an energy gap
and gapless edge states.
As a result, the system offers opportunities to explore magnetoplasmons
and edge magnetoplasmons.
Below, we will consider some examples to illustrate these points.

The group velocities of both the bulk and edge magnetoplasmons are 
given by Eq.~(\ref{eq:gvelqhe}),
which can be expressed in terms of the filling factor as:
\begin{align}
 v_\nu = \frac{n}{eB_{az}} \frac{e^2}{4\epsilon_0} = \frac{1}{4\epsilon_0} \frac{\nu e^2}{h}.
\end{align}
The propagation velocity is determined solely by the quantized Hall conductance 
$\nu e^2/h$ and permittivity.
The bulk magnetoplasmon is not specific to QHE.
For example, we expect it to be observable at $\nu=1/2$,
where the longitudinal resistance does not vanish.
The dispersion of the bulk magnetoplasmon at $\nu=1/2$ is given by Eq.~(\ref{eq:Dmpsimp}) as follows:
\begin{align}
 \omega_{mp}^{\nu=1/2} = \frac{eB_{az}}{m} + \frac{1}{4\epsilon_0} \frac{e^2}{2h} k_x.
\end{align}
As previously mentioned, there are multiple candidates for the ground state,
and there is a possibility that different surface plasmon quanta exist depending on the candidate.
Let us explore this possibility by adopting composite Fermion picture
for $\nu=1/2$.~\cite{Jain1989,Willett1993}
In this picture, the ground state $|{\rm QH}(\nu)\rangle$ has a nontrivial structure:
the external magnetic field is effectively ``erased'' by flux attachment to the electrons.
As a result, the effective mass changes from $m$ to $m^*$,
and we expect that a surface plasmon (in the absence of an external magnetic field)
made of composite fermions will appear. 
Assuming that all the electrons behave as composite fermions, 
the dispersion is given by Eq.~(\ref{eq:omegap}),
\begin{align}
 \omega_p^{\nu=1/2} =\sqrt{\frac{n e^2 k_x}{2 m^* \epsilon_0}} = \sqrt{ \frac{\omega_c^*}{2\epsilon_0} \frac{e^2}{2h} k_x}.
\end{align}
Since the two modes, $\omega_{mp}^{\nu=1/2}$ and $\omega_p^{\nu=1/2}$,
have different frequencies and group velocities, 
it would be possible to distinguish which mode is observed.

The variety of ground states that can arise in the fractional QHE
is notable, especially since electron-hole symmetry is absent ($1-\nu\ne \nu$).
Furthermore, even for a fixed $\nu$, 
the ground state can depend on the strength of $B_{az}$, i.e., 
the ground state becomes a function of $B_{az}$ and is denoted as $|{\rm QH}(\nu,B_{az}) \rangle$.
Let's take $\nu=2/3$ as an example.
In this case, we can consider the electrons as the carriers with $\nu_e=2/3$, 
while the holes can be treated as the mobile carriers, with $\nu_h=1/3$, 
in the filled lowest Landau level of $\nu_e=1$.
The ground states of these states are distinct, 
and two magnetoplasmons having different group velocities 
can be expected to exist:
\begin{align}
 & \omega_{mp}^{\nu_e=2/3} = \frac{eB_{az}}{m} + \frac{1}{4\epsilon_0} \frac{2e^2}{3h} k_x, \\
 & \omega_{mp}^{\nu_h=1/3} = \frac{eB_{az}}{m_h(B_{az})} + \frac{1}{4\epsilon_0} \frac{e^2}{3h} k_x.
\end{align}
The fractional QHE arises from the mutual interaction between electrons, 
which opens an energy gap between the ground state and the excited states.
This gap differs from the Landau level spacing, and $m_h$ is induced by the Coulomb interaction.
As a result, $m_h$ differs from the band mass $m$, and it can depend on $B_{az}$.
Therefore, the two dispersion curves can intersect at some $B_{az}$,
and away from this intersection, 
one of them may dominate at either lower or higher $B_{az}$. 
This may be observed as a change in the group velocity.~\cite{France2025}

Du {\it et al.} conducted a resonant inelastic light scattering experiment 
on various $\nu$ states within the second Landau level ($2\le \nu \le 3$)
and observed surface plasmons.~\cite{Du2019}
The square root dependence of the carrier density $n$ on $\omega_p$
in Eq.~(\ref{eq:omegap}) was clearly observed in an electron-hole symmetric 
manner:~\footnote{The external magnetic field has both perpendicular and parallel components 
with respect to the 2DEG. 
The tilted magnetic field plays a crucial role 
in ensuring that $k_x$ has a finite value in $\omega_p$.}
the dispersion reaches its maximum at $\nu=5/2$ and vanishes at $\nu=2$ and 3.
The spectrum, including both intensity and lifetime, was found to be 
rather informative as it reflected the quantum phases of the ground states.
For example, the intensity drops at $\nu=5/2$, suggesting the formation of the gaped fractional QH state.

%Magnetoplasmons are intrinsically related to magneto-rotons: 
%the latters are derived from the lowest Landau level projection of the Hilbert space, 
%while the former can be derived without this approximation.
The collective description of strongly interacting electrons in the fractional QHE 
was developed by Girvin, MacDonald, and Platzman (GMP).~\cite{MacDonald1985,Girvin1986}
They identified the branch of the magnetoplasmon dispersion relation
known as the magnetoroton, in analogy with the roton spectrum of superfluid helium.
Their projected density operator, $\bar{\rho}_{\bf k}$,
is constructed from the noncommutative guiding-center coordinates and obeys 
an infinite-dimensional algebra known as the W$_\infty$ algebra.~\cite{Ezawa2013}
The magnetoroton states are represented by $\bar{\rho}_{\bf k}|\Psi_{\rm L}\rangle$,
where $|\Psi_{\rm L}\rangle$ denotes the Laughlin state.
Haldane later proposed a geometrical description 
to explore the local collective degrees of freedom in a fractional QH state.~\cite{Haldane2011}
Observing that the quadratic combinations of guiding-center coordinates
generate area-preserving deformations,
he introduced an intrinsic metric for the fractional QH state, 
whose scalar curvature reflects local density fluctuations.
The excited state defined by $e^{i\gamma_{ab}\Lambda^{ab}}|\Psi_{\rm L}\rangle$, 
where $\gamma_{ab}$ is a symmetric, traceless shear tensor and 
$\Lambda^{ab}$ is the generator of area-preserving deformations,
can be viewed as a state formally analogous to a quantum squeezed state of the guiding-center geometry, 
representing fluctuations of the intrinsic metric.
In Haldane’s formulation, the fractional QHE is not merely a topological phase but also a quantum geometric fluid
in which geometry and density are dynamically intertwined.

The residual gauge freedom in the Coulomb gauge is governed 
by harmonic functions satisfying $\nabla^2 \lambda=0$. 
Interestingly, such harmonic gauge functions can generate 
area-preserving deformations on the surface.
For instance, $\lambda(x,y) \propto xy$ yields anisotropic dilation (hyperbolic scaling transformation)
$\xi^a = \epsilon^{ab} \partial_b \lambda$ with $\partial_a {\xi}^a = 0$,
where $\epsilon^{ab}$ denotes the two-dimensional antisymmetric tensor (Levi-Civita symbol). 
A symmetric shear mode, represented by the deformation $\lambda(x,y)\propto x^2-y^2$,
provides an example of an infinitesimally area-preserving deformation.
This observation suggests that 
the residual gauge transformations of evanescent electromagnetic fields 
may provide a physical realization of Haldane’s geometric description, 
in which local metric deformations correspond to area-preserving (active) diffeomorphisms:
$e^{i \int \lambda({\bf r}) \rho({\bf r}) d{\bf r}}|\Psi_{\rm L}\rangle$.
Exploring this correspondence could lead to a unified framework 
in which finite-lifetime surface-plasmon modes are interpreted 
as geometric collective excitations emerging from the residual gauge degrees of freedom.

As an effective field theory of the QHE, 
the Chern-Simons theory plays a central role.
This theory preserves gauge symmetry in the bulk; 
however, at the boundary, gauge transformations generate surface terms, 
making the action not strictly gauge invariant.
Wen showed that, in order to restore this invariance, 
a one-dimensional chiral boson theory naturally emerges at the edge.~\cite{Wen1990a}
Together, the bulk and edge form a gauge-invariant total system.
In other words, 
the chiral boson represents a promotion of 
the boundary gauge degrees of freedom to physical degrees of freedom, 
and the cancellation of gauge variations by the chiral boson is essential 
for the consistency of the bulk theory.
In this paper, we have shown that a similar structure appears
in the quantum state of surface plasmons, 
which arise from the requirement of preserving residual gauge symmetry
imposed at the surface of a 2DEG.
This correspondence illustrates that, just as in the QHE, 
the requirement of gauge invariance gives rise to a specific 
physical excitation---in this case, the surface plasmon mode.

\section{Discussion and outlook}\label{sec:dis}

Residual gauge transformations should not be considered entirely redundant, 
as they can have physical consequences. 
In particular, they manifest in the Aharonov–Bohm effect, especially 
when the $y$-axis is periodic, such as in a cylindrical geometry.
In this case, $C_y$ represents the Aharonov-Bohm flux, 
which can modify $|vac\rangle$ to create an ``excited'' ground state $|exc \rangle$.
Consequently, $J_y$ can sustain a static vacuum current, $\langle exc| J_y | exc \rangle \ne 0$, 
known as the persistent current in a metal ring.~\cite{Buttiker1983,Bleszynski-Jayich2009}
Since this current generates a nonzero static magnetic field, 
the photonic ground state is not simply $|0 \rangle$ but rather a coherent state,
where $A_y$ acquires a vacuum expectation value.
A residual gauge transformation
effectively alters the wavevector of the electron according to the covariant derivative.
Besides, a change in the wavevector can also result from interactions between electrons and the lattice.
If the coupling constant is appropriately tuned based on the value of the wavevector 
(e.g., for valley degrees of freedom) 
to preserve time-reversal symmetry, $C_j$ can represent lattice vibrations or phonons. 
Consequently, our formulation, which relates residual gauge transformations 
to Joule heating, can be extended in multiple directions.
Indeed, the Joule heating operator exhibits a fundamental uncertainty relation with the gauge interaction, 
expressed as $[V,\hat{Q}]=-i(\hbar/\epsilon_0) \int J_i J_i dx$.

The TE mode has a real frequency component in the Lorentz model.
Similar to the Drude model under an external magnetic field, 
the TE mode in the Lorentz model can hybridize with the TM mode
at the boundary, resulting in an edge magnetoplasmon in semiconductors.
With the introduction of a new energy scale, $\omega_0$, 
in addition to $\omega_c$ and $\omega_p$, 
the physics of edge magnetoplasmons can exhibit greater diversity.

Suppose that
a single photon in free space with frequency $\omega_{in}$
is converted into a single quantum of the TM mode with frequency $\omega$ ($\ll \omega_{in}$).
If both quanta share the same wavevector $k_x$, how can energy conservation be satisfied?
There are two possible explanations.
One possibility is that the matter Hamiltonian compensates for the energy difference, i.e.,
$\langle {\rm TM}(k_x)|H_M |{\rm TM}(k_x)\rangle-\langle vac|H_M|vac \rangle= \omega_{in}-\omega$.
However, this condition is not satisfied because 
the matter states under consideration are mostly virtual states,
making the left-hand side nearly zero on average.
The other possibility is that a new photon with energy $\omega_{in}-\omega$ 
is generated and returns to free space.
This explanation is more plausible, 
suggesting that when discussing the generation mechanism of these quanta,
we must necessarily consider light emission.
In this context, the residual gauge transformation must
acquire new physical significance.
Soon after the generation, the TM mode with $k_x$ arrives at the boundary.
It can be scattered back into the bulk region of the 2DEG due to the momentum-energy mismatch.
If the boundary width is reduced,
lateral confinement may cancel the localization perpendicular to the plane, i.e., $k_y \sim \xi^{-1}$.
In the dispersion relation,
\begin{align}
 \frac{\omega^2}{c^2} - k_x^2 - k_y^2 + \frac{1}{\xi^2} = 0,
\end{align}
this cancellation restores the light dispersion relation $\omega=ck_x$,
leading to a smooth transition from the TM mode to light at the boundary.
For surface plasmons, 
the frequency may be a more flexible entity (depending on the geometry of 2DEG)
than the light frequency in free space, 
and the distinction between off-shell and on-shell characters may become ambiguous.
%\begin{align}
% \frac{\omega}{c} = \frac{k_x}{\cosh\theta}, \ \ 
% \frac{1}{\xi} = k_x \tanh\theta.
%\end{align}
%Here, $\omega$ is taken as a constant specific frequency of the plasmon 
%that arrives at the boundary.
%When the outside of the boundary is simulated by $\theta=\infty$,

%When the outside of the boundary is simulated by $\theta=0$,
%the momentum can match to that of radiation while $\xi$ needs to be huge.
%This represents that the after passing the boundary the plasmon leaves as radiation.
%Thus, $\theta$ can represents a mixing between the quanta of surface plasmons
%and radiation.

%It is possible to apply the results to interband transitions.
%We consider graphene as an example.
%$\sigma_{xx} = \pi \alpha \epsilon_0 c$.
%$\omega=-i/\tau$, where $\tau_m = (2/\pi \alpha) (\xi/c)$.

\section*{Acknowledgments}

I am deeply grateful to Prof. Y. Tokura for helpful discussions and for his encouragement of this line of research.

%\appendix
%
%
%
%\section{The zero mode}\label{app:zero}
%
%
%A residual gauge transformation inludes the extending wave (not the evancenet waves)
%characterized by the wave vector perpendicular to the plane and angular frequency
%as
%\begin{align}
% & A^R_x(z,t) = \sum_{k_z} \left( e^{ik_z(z-ct)} a_R(k_z) + e^{-ik_z(z-ct)} a^\dagger_R(k_z) \right), \\
% & A^L_x(z,t) = \sum_{k_z} \left( e^{ik_z(z+ct)} a_L(k_z) + e^{-ik_z(z+ct)} a^\dagger_L(k_z) \right), 
%\end{align}

%\bibliographystyle{apsrev4-1}
%\bibliographystyle{apsrev}
% \bibliography{/Users/Sasaki/tex/bib/library}
%\bibliography{/Users/sasakikenichi/bib/sasaki,/Users/sasakikenichi/bib/library.bib}


\end{document}